\documentclass[12pt]{scrartcl} 
\usepackage[usenames,dvipsnames]{xcolor}

\usepackage[colorlinks]{hyperref}
\hypersetup{citecolor=MidnightBlue,linkcolor=BrickRed,urlcolor=OliveGreen}

\usepackage{amsmath,amssymb}
\usepackage{todonotes}
\usepackage[round]{natbib}
\usepackage{pdfpages}
\usepackage{subcaption}
\usepackage{booktabs}
\usepackage{authblk}
\usepackage{bm}

\usepackage[page]{appendix}
\providecommand{\keywords}[1]{\textbf{\textit{Key-words ---}} #1}
\usepackage{setspace}
\usepackage{caption}
\title{Scaling multi-species occupancy models to large citizen science
datasets}
\author[1, *]{Martin Ingram}
\author[2, 3]{Damjan Vukcevic}
\author[4]{Nick Golding}
\affil[1]{School of BioSciences, University of Melbourne, Parkville, VIC 3010,
  Australia}
\affil[2]{School of Mathematics and Statistics, University of Melbourne,
  Parkville, VIC 3010, Australia}
\affil[3]{Melbourne Integrative Genomics, University of Melbourne, Parkville,
  VIC 3010, Australia}
\affil[4]{Telethon Kids Institute and Curtin University, Perth, Western
  Australia, Australia}
\affil[*]{Corresponding author: Martin Ingram, ingramm@student.unimelb.edu.au}

\date{}

\begin{document} 
\maketitle
  \begin{enumerate}
  \item Citizen science datasets can be very large and promise to improve
    species distribution modelling, but detection is imperfect, risking bias
    when fitting models. In particular, observers may not detect species that
    are actually present. Occupancy detection models can estimate and correct
    for this observation process, and multi-species occupancy detection models
    exploit similarities in the observation process, which can improve estimates
    for rare species. However, the computational methods currently used to fit
    these models do not scale to large datasets.

  \item We develop approximate Bayesian inference methods and use graphics
    processing units (GPUs) to scale multi-species occupancy detection models to
    very large citizen science data. We fit multi-species occupancy detection
    models to one month of data from the eBird project consisting of 186,811
    checklist records comprising 430 bird species. We evaluate the predictions
    on a spatially separated test set of 59,338 records, comparing two different
    inference methods -- Markov chain Monte Carlo (MCMC) and variational
    inference (VI) -- to occupancy detection models fitted to each species
    separately using maximum likelihood.

  \item We fitted models to the entire dataset using VI, and up to 32,000
    records with MCMC. Variational inference fitted to the entire dataset
    performed best, outperforming single-species models on both AUC (90.4\%
    compared to 88.7\%) and on log likelihood (-0.080 compared to -0.085). We
    also evaluate how well range maps predicted by the model agree with expert
    maps. We find that modelling the detection process greatly improves
    agreement and that the resulting range maps agree as closely with expert
    maps as ones estimated using high quality survey data.

  \item Our results demonstrate that multi-species occupancy detection models
    are a compelling approach to model large citizen science datasets, and that,
    once the observation process is taken into account, they can model species
    distributions accurately.

  \end{enumerate}


\section{Introduction}

Citizen science datasets, that is, datasets collected by volunteers, can be extremely useful for ecological research. Their value lies in their large spatial and temporal scale: datasets are often collected across large areas, and throughout the year. This allows ecologists to investigate important questions that would be hard or impossible to answer with conventional datasets. Such questions are, for example, how species ranges and migration patterns are shifting over time \citep{citizen-science-research-tool}. 

Aided by the use of the internet and smartphones, modern citizen science projects often amass enormous amounts of data. In this work, we focus on the eBird dataset, a dataset of bird sightings. It contains over 100 million bird sightings each year\footnote{see \url{https://ebird.org/about}}, collected throughout the year and across the world \citep{ebird-2009}. This makes it far larger than many other datasets. For example, the North American Breeding Bird Survey (BBS) \citep{pardieck2020north} is collected only once a year and consists of a comparatively small number of sightings. 

Datasets as large and frequently collected as eBird should allow ecologists to investigate more hypotheses. For example, the larger sample of eBird may allow estimates of ranges for species that are rarely observed in the BBS. However, analysing such datasets is complicated by the quality of the sightings. While the BBS is a high-quality survey, with a standardised protocol and skilled observers, eBird sightings are made following different protocols, by observers with different skill levels, at different times of day. As a result, while the BBS records can plausibly be treated as presences or absences along sampling routes (as done, for example in \cite{mistnet}), doing the same for eBird data risks introducing severe bias.

The data could also be modelled as presence-only data, another well-established
approach in species distribution modelling. However, this fails to make use of a
crucial part of eBird data: participants are asked whether they reported all
species they were able to identify confidently. Participants are encouraged to do so, and in the majority of records, all species are reported. In this case, the absence of a
record implies that the species was not observed, and thus that the data is more
than just ``presence-only''. This is crucial, since only limited inferences can
be drawn from presence-only data. At best, relative suitability of sites can be
assessed (see \cite{fit-for-purpose}, for example). We note that eBird also contains some ad-hoc sightings where participants may report only a subset of species, but we focus on the complete records in this paper.

\emph{Occupancy detection models} are a natural approach to model datasets like
eBird \citep{mackenzie-occ-det-2002,altwegg-2019}. These models assume
that, at each visit to a site, a species is detected with probability $p$ if it
is present; this probability of detection is a function of covariates, such as
the time of day and time spent birding. The probability of presence $\Psi$ at
the site is inferred jointly with $p$; it is modelled as a function of
environmental covariates. In this way, occupancy detection models aim to
separate the biological process determining $\Psi$ from the observation process
determining $p$.

Occupancy detection models can be fitted separately to each species by maximum
likelihood \citep{mackenzie-occ-det-2002}. A \emph{multi-species} occupancy
detection (MSOD) model instead models all species at once by placing a
hierarchical Bayesian prior on the coefficients modelling the observation
process $p$\footnote{A hierarchical prior can also be placed on the coefficients
  modelling the biological process $\Psi$, but we focus on the observation
  process here.}. This is particularly useful when modelling rare species: for
these species, the detection coefficients will be shrunk towards the group mean,
which is often reasonable, since we expect similarity in the observation
processes across species. MSOD models were initially proposed by \cite{Kery2009}
and are typically fitted with Markov chain Monte Carlo (MCMC) software such as
\texttt{WinBUGS} or \texttt{JAGS}.

Both single-species and multi-species occupancy detection models are well known
in ecology, and the potential advantages of modelling imperfect detection for
species distribution modelling and citizen science data have been discussed in
previous work (see \cite{arroita-2017,altwegg-2019}, for
example). However, existing methods for fitting MSOD models are unable to scale
to large citizen science datasets such as eBird. For example, recent work
\cite{arroita-occ-det2019} reported that it took hours to fit an MSOD model to
around 1,000 observations of about 100 species. We will see that even a single
month of eBird data consists of almost 200,000 checklists, suggesting that a
different approach is needed to fit models to this data in a reasonable amount
of time.

We make the following contributions in this paper. Firstly, we develop efficient
Bayesian inference approaches which allow us to fit MSOD models to very large
datasets. To do so, we propose a sparse data structure to make evaluation of the
likelihood more efficient and make use of Graphics Processing Units (GPUs) to
accelerate computations. We fit models using both MCMC, and variational
inference. Secondly, we evaluate the predictive performance of these models on a
spatially separated test set. We find that they outperform maximum likelihood
fitted separately to each species on both AUC and log likelihood, especially for
rarely-observed species. The variational inference approach performed best: it
fitted the full dataset in under two hours while achieving similar evaluation
metrics as MCMC, which took multiple days to fit. Finally, we demonstrate that
predictions made by the model agree as well with expert range maps from BirdLife
International \citep{birdlife-maps} as those estimated using data from the North
American Breeding Bird Survey \citep{pardieck2020north}. Compared to the BBS, we argue that using eBird data has two main advantages: (1) over 80\% of species observed fewer than eight times in the BBS are observed at least 20 times in the eBird dataset, expanding the number of species for which we can produce reliable range maps, and (2) eBird data are collected throughout the year. We thus believe that our approach is compelling for large citizen science datasets, and are excited about the opportunities it offers for advancing ecological research.

\section{Methods}
\subsection{Model}
\label{subsec:model}

In this section, we briefly revisit the assumptions behind multi-species
occupancy detection modelling. The true presence or absence of a species $j$ at
site $i$, $y_{ij}$, is modelled conditional on a column vector of environmental
covariates $\bm{x}^{\text{(env)}}_i$ for that site. This is the same assumption
made in many presence/absence (PA) models. The difference is that PA models
assume that $y_{ij}$ is observed directly, while occupancy detection models do
not. Mathematically speaking, we model $y_{ij}$ as follows:
\begin{align}
y_{ij} \sim \textrm{Bern}(\Psi_{ij}), \\
  \textrm{logit}(\Psi_{ij}) = \bm{x}^{\text{(env)}\intercal}_i \bm{\beta}^{\text{(env)}}_j + \gamma_j, \\
  \bm{\beta}^{\text{(env)}}_j \stackrel{iid}{\sim} \mathcal{N}(0, I), \\
  \gamma_j \stackrel{iid}{\sim} \mathcal{N}(0, 10^2).
\end{align}
Here, the vector of coefficients $\bm{\beta}^{\text{(env)}}_j$ models the environmental
response for species $j$, and $\gamma_j$ is a species-specific intercept.

Occupancy detection models assume that there is a possibility that any given
observer may have overlooked a species that is actually present. Observers may
also incorrectly identify a species, but in this study we make the common
assumption that there are no false positives. We refer the interested reader to
\cite{altwegg-2019} for a detailed discussion of the assumptions made in
occupancy detection modelling.

Our dataset takes the form of records $s_{ijk}$, where $s_{ijk} = 1$ if the
$k$-th observation for species $j$ at site $i$ was a presence, and zero
otherwise. We make the following assumptions for $s_{ijk}$:
\begin{align}
  & p(s_{ijk} = 1 \mid y_{ij} = 1) = p_{ijk}, \label{eq:det-prob} \\
  & p(s_{ijk} = 1 \mid y_{ij} = 0) = 0, \label{eq:false-pos} \\
  & \textrm{logit}(p_{ijk}) = \bm{x}_{ik}^{\text{(obs)}\intercal} \bm{\beta}^{\text{(obs)}}_j \label{eq:det-prob-model}.
\end{align}
In words, \autoref{eq:det-prob} says that the probability of observing species
$j$ in site $i$ on the $k$-th visit, assuming that it is present, is $p_{ijk}$;
\autoref{eq:false-pos} rules out false positives; and
\autoref{eq:det-prob-model} sets up a logistic regression model for the
detection probability in terms of the observation covariates.

The key feature of the \emph{multi-species} occupancy detection models used in
this paper is that the species-specific observation coefficients
$\bm{\beta}_j^{\text{(obs)}}$ are given a hierarchical prior:
\begin{align}
  \beta_{jl}^{\text{(obs)}} &\stackrel{iid}{\sim} \mathcal{N}(\mu_l, \sigma_l^2), \\
  \mu_l &\stackrel{iid}{\sim} \mathcal{N}(0, 1), \\
  \sigma_l &\stackrel{iid}{\sim} \mathcal{N}^+(1),
\end{align}
where $\mathcal{N}^+(1)$ denotes the half-normal distribution with a standard
deviation of $1$. This prior models each species' coefficient as drawn from a
group prior distribution. As mentioned in the introduction, this is useful since
we expect that many observation coefficients across species are likely to be
similar. For example, for most species, the probability of observation is likely
to be lower at night, so the group mean $\mu_l$ for a ``night-time'' indicator
would likely be negative, and its variance $\sigma_l^2$ would indicate the
strength of this prior belief.

We note that the vector of environmental coefficients
$\bm{\beta}_j^{\text{(env)}}$ could similarly be given a hierarchical prior, and
this is done in many MSOD models. When designing the model, we chose not to do
this as niches can differ greatly between species and thus we expected that
pooling coefficients towards a group prior would be less useful than for the
observation process, but the extension would be straightforward to implement.

The assumptions above lead to the following likelihood for all $N$ sites, $J$
species, and $K = \sum_{i} K_i$ checklists, where $K_i$ is the number of
checklists at site $i$:
\begin{align}
  p(y, s \mid \theta) = \prod_{i=1}^{N} \prod_{j=1}^J p(y_{ij} \mid x_i) \prod_{k=1}^{K_i} p(s_{ijk} \mid y_{ij}, x_{ik}^{\text{(obs)}}).
\end{align}

This is the complete-data likelihood. In practice, only $s$ is observed. To
produce a likelihood for $s$ only, $y$ can be summed out (see section A of the
supplementary material) to yield:
\begin{align}
  p(s \mid \theta) = \prod_{i=1}^{N} \prod_{j=1}^{J} \left[(1 - \Psi_{ij}) \prod_{k=1}^{K_i} (1 - s_{ijk}) +
  \Psi_{ij} \prod_{k=1}^{K_i} (p_{ijk})^{s_{ijk}} (1 - p_{ijk})^{1 - s_{ijk}} \right].
  \label{eq:observed-data-likelihood}
\end{align}

\subsection{Computational approaches}

The model presented in the previous subsection is well known. The likelihood in
\autoref{eq:observed-data-likelihood} was first proposed for a single species by
\cite{mackenzie-occ-det-2002}, where it was used to fit models with maximum
likelihood. The extension to multiple species with hierarchical priors on
coefficients was first proposed by \cite{Kery2009}. There, the authors fit the
model to 267 1km$^2$ units sampled 2-3 times for 134 species, focusing on the
estimation of species richness. The same dataset, this time considering 158
species, was analysed by \cite{arroita-occ-det2019}, where the authors wrote
that the model took about 7 hours to fit with the modelling software
\texttt{JAGS}. It is usually fitted using a data augmentation approach
originally proposed by \cite{royle-multinomial-2007}, which is more efficient
than directly using the likelihood in \autoref{eq:observed-data-likelihood} when
sampling using the MCMC algorithms implemented in \texttt{WinBUGS}
\citep{Lunn2000}. While these approaches are well understood and trusted, as
discussed previously, they are unable to fit datasets as large as eBird in a
reasonable time.

To accelerate computations, we implemented the code to evaluate the log
posterior density using Google's \texttt{JAX} package \citep{jax2018github},
which automatically parallelises code evaluation and is able to use GPUs. Rather
than using the augmentation approach of \cite{royle-multinomial-2007}, we
calculate the likelihood as it is stated in
\autoref{eq:observed-data-likelihood}, which we found to be efficient. This
means we can do without the additional assumptions made by
\cite{royle-multinomial-2007} and avoid introducing latent parameters,
simplifying the model's interpretation.

We use two inference methods: MCMC using the No-U-Turn sampler (NUTS)
\citep{hoffman14a}, and mean-field variational inference. For MCMC, we used NUTS
as implemented in the \texttt{numpyro} software package
\citep{phan2019composable}. NUTS is the main approach used by many well-known
software packages, such as \texttt{Stan} \citep{rstan}. It is a version of
Hamiltonian Monte Carlo (HMC), used for example in the \texttt{greta} package
\citep{Golding2019}. NUTS and HMC are often more efficient than the MCMC
algorithms used in \texttt{WinBUGS} and \texttt{JAGS} \citep{faster-bayes-2016}.

Our mean-field variational inference approach is based on automatic
differentiation variational inference (ADVI) \citep{advi-paper}, specifically
the variant proposed by \cite{giordano-robust}. This variant allows the use of
second-order optimisation, rather than the stochastic optimisation used for
ADVI. We prefer second-order optimisation for two reasons: firstly, it has no
tuning parameters, and secondly, convergence is easily detected. We implemented
the approach in \texttt{JAX}. We provide both a python package and an R package
to allow users to fit our approaches.\footnote{The python package is available
  at \url{https://github.com/martiningram/occu_py}, and the R package
  (recommended) is available at \url{https://github.com/martiningram/roccupy}.}
Unlike MCMC, variational inference does not provide samples from the exact
posterior distribution. Instead, it approximates the posterior distribution
using independent univariate normal distributions, one for each parameter. The
parameters of these distributions are optimised to minimise the Kullback-Leibler
divergence between the approximation and the posterior distribution. This type
of variational inference tends to underestimate parameters' marginal variances
and is also unable to capture posterior correlations. However, it often provides
good estimates of the posterior means, and it can be much faster to fit. Later
in this paper, we find that when predicting to unseen data, the gap between MCMC
and ADVI narrows as more data is used, suggesting that modelling correlations
becomes less important when data are abundant. We include more mathematical
details about ADVI in section D of the supplementary material. Our version of
ADVI has one parameter that has to be chosen: the number of fixed draws, $M$,
used to evaluate an expectation in the objective. A larger $M$ yields more
precise estimates of the objective, so it should be set as large as possible
given the available memory constraints.

As mentioned in the introduction, occupancy detection models are commonly
estimated using maximum likelihood fitted to each species separately. To compare
our methods against this approach, we had initially planned to use the
implementation in the software package \texttt{unmarked} \citep{unmarked} but
found it to be very inefficient. We believe this is because of two
factors. First, \texttt{unmarked} does not use gradient information to optimise
the objective, which requires using costly numerical differentiation
instead. Second, and more importantly, \texttt{unmarked} stores the observations
as a matrix of size $N \times K_\text{max}$, where $K_\text{max}$ is the maximum
number of observations at a site. If a site is visited fewer than $K_\text{max}$
times, the remaining observations are coded as missing. This is convenient when
sites are visited roughly equally often, but becomes very inefficient when they
are not. In the dataset used in this analysis, most sites are visited just once,
but some are visited hundreds of times, resulting in a matrix of observations
that contains a very large number of missing entries.

To avoid this inefficiency, we instead encode observations using one long binary
vector for each species, together with another vector of integers specifying
which site each observation was made in. We use this sparse representation to
efficiently estimate the likelihood both for the hierarchical models proposed
and for a faster implementation of maximum likelihood. Our focus in this paper
is on hierarchical occupancy detection models, but we show in the supplementary
material that the fast maximum likelihood implementation recovers the same
coefficients as \texttt{unmarked} on an example dataset (section G), and that it
scales much more efficiently as the number of checklists is increased (section
B).

\subsection{Dataset}

We used the eBird Basic Dataset (EBD) for our analysis. We focused on a single
month, June 2019, and limited records to the United States of America. We also
limited the species in the analysis to those found in the North American
Breeding Bird Survey to facilitate comparisons with that dataset. Records
outside of the mainland of the US were dropped. This resulted in 430 species
with at least one observation. We further used the \texttt{blockCV} software
package \citep{valavi2019block} to split the dataset into four folds, of which
we used three as a training set and the fourth as a spatially separated test
set.

The basic unit of observation in the eBird dataset is the checklist. When an
eBird user goes bird watching, they are asked to record the species they were
able to identify. An important difference between eBird data and other citizen
science datasets is that participants are asked whether they reported all
species they were able to identify on their trip, or whether they included only
a subset. They are strongly encouraged to report all species they were able to
identify, and generally do so. We used only these complete records. The absence
of this flag would complicate modelling, as it would be unclear whether an
observer truly did not observe a species or chose not to report it.

In addition, eBird users are required to record a number of additional fields,
depending on the protocol used. We subsetted the data to the three most common
protocols:
\begin{enumerate}
\item \emph{Traveling} and \emph{Traveling -- property specific} (58.4\% of
  checklists): Birds observed while travelling. Observers record the duration of
  their trip as well as the distance travelled.
\item \emph{Stationary} (40.0\% of checklists): Birds seen while the observer
  stays at a certain location, recording the time spent.
\item \emph{Area} (0.4\% of checklists): Birds seen while the observer
  systematically searches an area. Both the time spent searching and the area
  searched are recorded.
\end{enumerate}

Each checklist record includes a single latitude and longitude entry. To ensure
that these spatial coordinates were close to the location of the observations,
we dropped checklists following the \emph{Traveling} protocol with distances
travelled greater than 3km, which resulted in roughly every third checklist in
this protocol being dropped. For the same reason, we dropped checklists in the
\emph{Area} protocol with areas of over 9~km$^2$ searched, of which there were
only five.

\begin{figure}
  \includegraphics[width=1\textwidth]{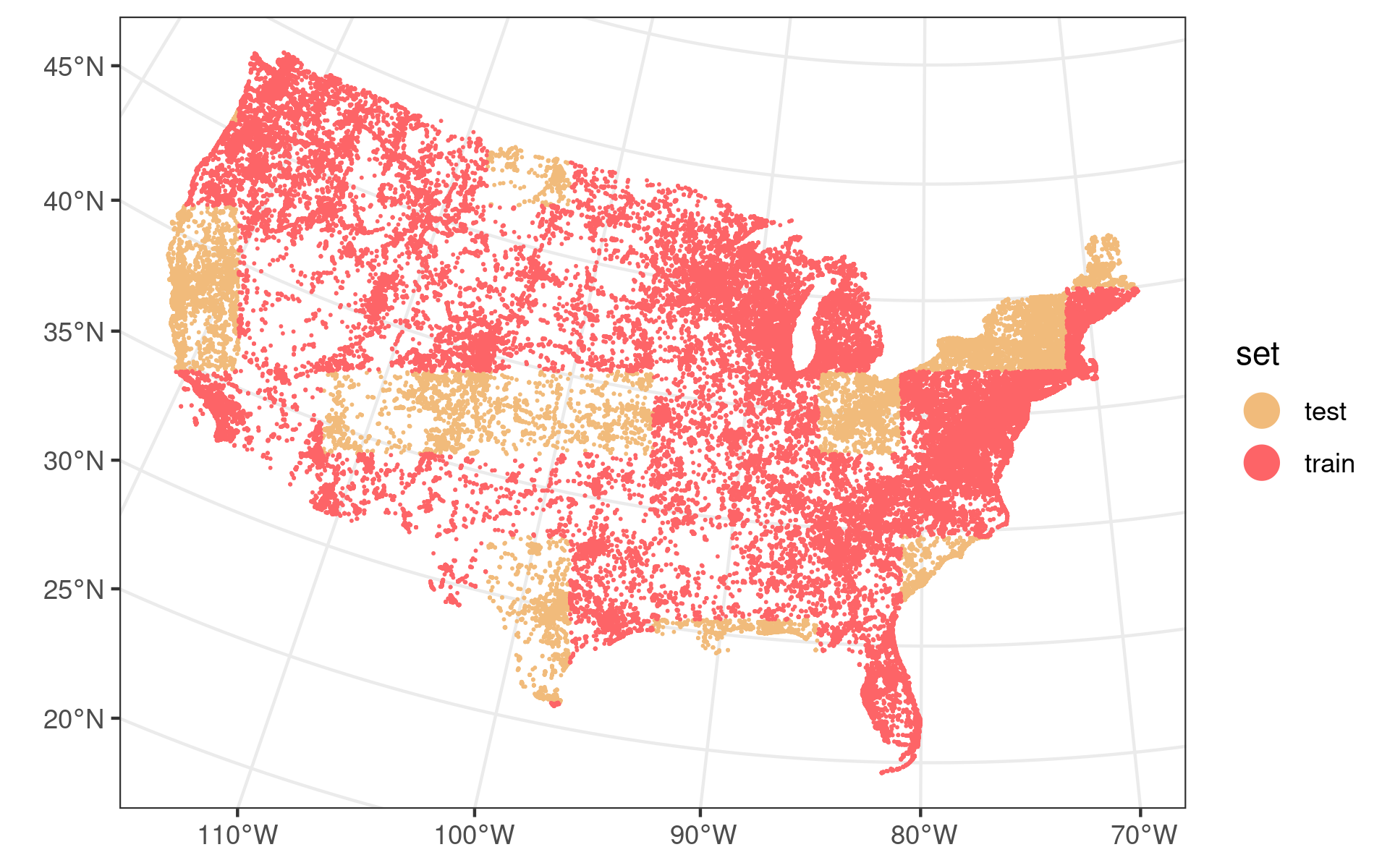}
  \caption{The locations of the checklists in the training set (red) and the
    test set (brown).}
  \label{fig:train-test-split}
\end{figure}

The full dataset consisted of 323,299 checklists. After all filtering steps, a
total of 246,149 checklists remained, of which 186,811 were in the training set,
and the remaining 59,338 formed the test set. The locations of these checklists
and the split into train and test sets chosen automatically by the
\texttt{blockCV} package are shown in \autoref{fig:train-test-split}.

Finally, we kept only species with at least five observations in the training
set, resulting in a total of 426. Even after these processing steps, the dataset
remained very large, at 186,811 observations of these 426 species. For
comparison, the Swiss bird dataset used in \cite{arroita-occ-det2019} consisted
of 267 sites sampled 2-3 times with records for 158 species. If every site were
sampled three times, this would result in 861 checklists. The full dataset
considered here thus contains over 100 times as many observations of almost
three times as many species, demonstrating the need for more scalable inference
techniques.

\subsection{Choice of covariates and sites}

To fit the model described in \autoref{subsec:model} to this dataset, we had to
choose which environmental variables $\bm{X}^{\text{(env)}}$ to include. Since
climate is likely to strongly influence bird distribution, we included the
worldclim climate variables \citep{hijmans2005very} at a resolution of 10
minutes of a degree, which is roughly equivalent to about $18 \times 18$~km$^2$
in area.\footnote{This is the area spanned at the equator; at the latitudes in
  question, the equivalent area will be somewhat smaller.} Among these, we
removed highly correlated covariates as follows: going through each column in
order, if a column of the design matrix had a sample correlation greater than
95\% with a previous column, we dropped it from the analysis. This eliminated
three of the nineteen covariates.

Occupancy detection models assume that the data consist of sites that are
repeatedly visited. In surveys, sites are chosen and visited several times,
while in eBird, checklist locations are not chosen systematically. To still be
able to apply occupancy detection models, we broke the area of interest into
sites. To do this, we used the environmental covariates to define a grid across
the study area. The worldclim covariates are constant within cells of roughly
$18 \times 18$~km$^2$ area, as described in the previous paragraph. Using these
cells as the sites $i = 1, ..., N$ resulted in a total of 15,011 sites
containing at least one checklist. Most cells (3,498) contained only a single
checklist, but some contained hundreds, the most being 928.

At this point, we would like to point out a potential issue with our choice of
data. The likelihood given in \autoref{eq:observed-data-likelihood} assumes that
sites are independent conditional on the observation and environmental
covariates. Although the assumption that checklists are collected independently
from each other is likely to be reasonable, the occupancy of cells is likely to
be spatially correlated, for example due to unmeasured covariates. Put simply,
knowing that a species is present in one cell likely raises the probability of
it being present in a neighbouring cell, even if the two locations have the
same, or very similar, environmental covariates. The model presented here could
be extended to also model spatial correlation, as was proposed for example in
\cite{spatial-occ-det}.

However, we choose not to pursue this route here, for the following
reasons. Firstly, from a practical point of view, a spatial model would raise
difficult computational challenges. Secondly, adding a spatial effect is also
complicated by the problem of spatial confounding \citep{spatial-occ-det,
  spatial-confounding} which, summarised briefly, says that adding a spatial
effect can cause bias in the estimates of the coefficients of a regression
model. To address spatial confounding, the spatially-correlated errors are often
modified to explain only variation not explained by the environmental
covariates. In such a modified model, the estimated environmental coefficients
have the same posterior means, but larger variances (see Figure 1 in
\cite{spatial-occ-det}, for example). This means that the model presented should
produce good posterior mean estimates of the environmental coefficients, even if
their variance estimates are likely to be somewhat overconfident. In summary, we
argue that while a spatial extension would be interesting and useful future
work, the simpler approach presented here can still produce reasonable
predictions.

In addition to the worldclim climate variables, we included land cover
covariates derived from the National Land Cover Database in its 2016 iteration
\citep{homer2012national}. We aggregated these as follows. Four of the
covariates related to ``developed'' areas of varying intensity. We grouped these
together into a single category. Three of the categories -- ``Unknown'',
``Barren Land'', and ``Perennial Ice/Snow'' -- were very rarely assigned, so we
grouped them together into the category ``Other''. Finally, we combined the two
categories relating to wetlands into a single category.

We derived binary indicators of whether each cell contained a particular land
cover. For example, the ``has\_open\_water'' covariate takes the value of 1 if
the cell is partly classified as ``Open Water'', and zero otherwise. We
considered two alternative summaries: an indicator of the majority land cover
for each cell, and the fraction of the cell occupied by each land cover type. We
chose the binary indicator over the majority cover since it encodes more
information. Compared to the fraction covered, we believe the indicator to be a
more useful summary in a linear model. Our reasoning is that, taking a water
bird as an example, the availability of \emph{any} water is likely to
drastically increase the probability that the cell is suitable, but the
difference between, say, 20\% or 40\% of the cell being covered by water is
unlikely to increase it much further.

A list of all environmental covariates used together with their means and
standard deviations is given in Table 1 in the supplementary material. We
further included the square of each worldclim covariate since the unimodal
environmental responses implied by quadratic terms are more ecologically
plausible than straight line fits. Some authors have argued that species
response curves are skewed and thus quadratic terms are not sufficient
\citep{AUSTIN20071}. The model could be extended to include spline terms or
interactions to improve its fit, but for simplicity, we included only quadratic
terms here. All in all, we used a total of 43 environmental covariates. All
covariates apart from the indicators were standardised to have mean 0 and
standard deviation 1 in the training data.

The observation covariates should be chosen to provide information about
$p_{ijk}$, the probability that a species is observed given that it is
present. To do this, we included the (log-transformed) duration spent birding in
minutes, as well as the protocol type (Stationary, Travelling, or Area). The
detection probability is also likely to be affected by the time of day. To take
this into account, we obtained the times of sunrise and sunset for each
checklist and categorised the observation times into eight categories based on
the time to sunrise and from sunset (please see section E of the supplementary
material for a description of these categories).

We also expected the land cover to affect the probability of detection, for two
reasons. Firstly, land cover can make it more difficult to detect species
presence. For example, a densely wooded area may reduce visibility compared to
an open grassy area. Secondly, checklists only cover part of the site: sites are
around $18\times18$~km$^2$ in size, but checklists cover a route of 3~km at
most. So, even though a site may be partly covered by forest, making it suitable
for a forest-dwelling bird, the observations may not be made in that part of the
cell. To take these factors into account, we used the NLCD dataset at a finer
$1.5 \times 1.5$~km$^2$ resolution to determine the dominant land cover for each
checklist and included this indicator as a covariate. We aggregated the land
cover covariates more coarsely for the detection probability, to ``Forest'',
``Developed'', ``Water'', and ``Other'', since we expected these to affect
detection probabilities the most. All together, this resulted in 13 covariates
used to model the observation process.

\subsection{Evaluation}

Evaluating occupancy detection models is challenging, since no ground truth is
available for the true presence and absence of species. We therefore evaluate
the different approaches in two stages: first, by how well they are able to
predict the observations in the spatially separated test set. This evaluation
only assesses the model's ability to predict joint detection and suitability, so
its ability to separate the two is not evaluated. Our assumption is that better
evaluation performance on the separated test set translates to improved
estimates of occupancy, but we acknowledge that this is not necessarily so. For
this reason, we also include a comparison of predicted suitability maps with
expert range maps, which provides a qualitative evaluation of the model's
ability to separate the two processes.

To evaluate the utility of using more data, we fitted the proposed inference
approaches -- variational inference and MCMC -- to different dataset sizes. We
created smaller datasets from the full one by randomly selecting a sample of
checklists from the training set, starting with 1,000 and doubling them until
32,000 checklists are reached. For variational inference, we also fitted 64,000
checklists, 100,000 checklists and, finally, the full training set of
186,811. As mentioned previously, the variational inference approach requires
fixing a number of draws $M$ to estimate the objective, with large $M$ reducing
bias but also requiring more memory. We used $M=100$ up to 4,000 checklists,
$M=50$ for 8,000, $M=25$ for 16,000 up to 100,000, and $M=15$ for the full
186,811 checklists. Using $M=100$ should incur only a small amount of bias, so
we did not explore using larger $M$ than this for the smaller number of
checklists. For the larger datasets, we chose $M$ to be as large as possible
given the available GPU memory.

We added two baseline models to compare against. First, we fitted the smallest
dataset of 1,000 checklists with \texttt{Stan} \citep{rstan}. We did this to
ensure that the MCMC approach in \texttt{numpyro} worked as expected. Second, we
fitted each species separately using maximum likelihood on the full dataset. We
used only the full dataset since, unlike the Bayesian models which fall back to
the prior in the absence of data, maximum likelihood inference requires a
minimum number of observations to avoid singular estimates. Some species are
observed very rarely, so only using the full dataset ensured that each species
was observed at least once in the data.

As we increased the number of checklists, we faced two computational issues.
When the dataset included more than 16,000 checklists, the MCMC approach in
\texttt{numpyro} failed to converge using its default settings. We were able to
address this by switching from single precision arithmetic to double
precision. When using the variational inference approach, the largest checklist
datasets required a large amount of GPU memory. For this reason, we used an
NVIDIA P100 GPU with 16GB of memory for the variational inference approaches,
while an NVIDIA GTX-2070 GPU with 8GB of memory was used for the MCMC approach.

To assess the quality of predictions, we use the area under the receiver
operating characteristic curve (AUC) and mean log likelihood. AUC estimates the
probability that a randomly chosen positive example will be ranked more highly
than a randomly chosen negative example \citep{fielding1997review}, making it
easy to interpret. However, AUC only assesses the ability of the model to rank
its predictions, not the calibration of the predicted probabilities. The log
likelihood, our second metric, also assesses calibration, and it is maximised
when the predicted probabilities are equal to the true probabilities, making it
a proper scoring rule \citep{gneiting-scoring}. We calculated each species' mean
log likelihood by computing the log probability assigned to the observed outcome
in each checklist, and then computing its average. We refer the interested
reader to \cite{lawson2014prevalence} for a detailed discussion of the
differences between AUC and log likelihood.

The metrics above evaluate how well the different methods are able to predict
unseen, spatially separated data from the eBird dataset. As mentioned at the
start of the section however, if model scores highly on these metrics, this does
not necessarily mean that it is a good model of species distribution. Each data
point in eBird is affected not only by the biological process of interest but
also the observation process, and it cannot be used to evaluate either in
isolation. To address this, we initially considered using the North American
Breeding Bird Survey (BBS) \citep{pardieck2020north}, a high quality survey
dataset, as ground truth. This dataset consists of routes, each of which is
divided into fifty stops, spaced half a mile apart. At each stop, surveyors
remain stationary for three minutes and record all bird species they were able
to detect. Participants in the survey are highly skilled, but we believe that
the BBS protocol is still likely to miss species that are actually present. For
example, observations are made by roads, which may make it hard to observe
certain kinds of species. They also tend to be made at a particular time of day
(usually in the early morning) and the time spent per stop is short.

In view of these issues, we instead compared the models' predictions against
expert range maps from \cite{birdlife-maps}. These expert maps also do not
represent ground truth: they are based on subjective judgement, and they are
generally not specified at a high level of detail. Still, good agreement with
expert maps should indicate that a model is in line with established knowledge,
which may build trust.

Our aim in using the expert maps was to answer two questions. First, how does
the best MSOD model fitted to eBird compare to the same model fitted to
high-quality BBS data; and second, how much is agreement improved when using an
MSOD model compared to treating the data as presence/absence. To answer the
first question, we fitted an MSOD model to BBS by treating the fifty stops made
along each survey route as repeat visits to the same site. For the second, we
fitted PA models to BBS and eBird. To do this for the BBS dataset, we treated a
species as present if it was observed at least once on a route and absent
otherwise. For eBird, we fitted a PA model by assuming that each checklist
consisted of presence/absence data. To keep the PA models as similar to the MSOD
models as possible, we used the same model for $\Psi$ as in
\autoref{subsec:model}, but treated $y$ as being directly observed.

To evaluate how well a model agrees with an expert map, we first predicted the
probability of presence, $\Psi$, for each cell of the worldclim covariates in
the study area. We then computed the square error between the predicted
probability and the value in the expert map for each cell (either 1 or 0). We
summarised the overall error by taking the mean of these square errors, which is
also known as the Brier score.

\section{Results}

\subsection{Evaluation results}

\begin{figure}
  \includegraphics[width=\textwidth]{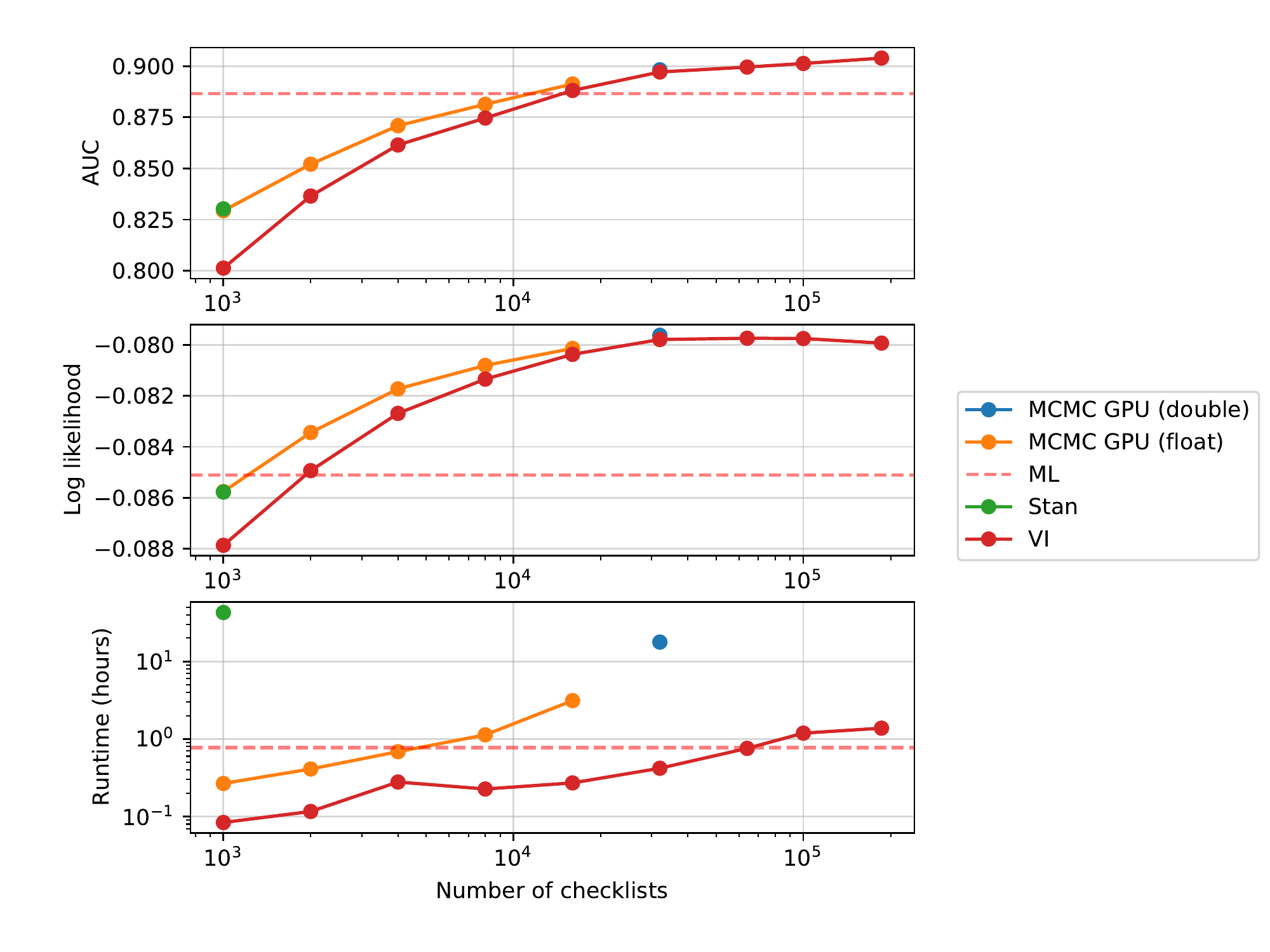}
  \caption{Evaluation results by number of checklists. The top panel shows AUC,
    the middle panel log likelihood, and the bottom the runtime in hours. Please
    note that the x-axis is on a log scale for all three panels, and the runtime
    is on a log scale also. The inference approaches are abbreviated as follows:
    VI denotes the multi-species occupancy detection model fitted with
    variational inference; MCMC GPU denotes the same model fitted using MCMC,
    with (float) denoting single and (double) double precision arithmetic;
    \texttt{Stan} refers to the same model fitted using the \texttt{Stan}
    software; finally, ML refers to fitting separate models to each species
    using maximum likelihood. The maximum likelihood result fitted to the entire
    dataset is shown dashed to aid visual comparisons against the other
    results.}
  \label{fig:eval-results-size}
\end{figure}

\autoref{fig:eval-results-size} shows the evaluation results for different
numbers of checklists. The top panel, comparing AUC, shows a clear increasing
trend as the number of checklists increases. At 1,000 checklists, MCMC clearly
outperforms VI, at around 83\% AUC compared to about 80.1\% for VI. This
difference shrinks as the dataset size is increased, however, and at 32,000
checklists, VI's AUC is essentially identical to MCMC's. This suggests that, for
prediction, the ability to capture posterior correlations becomes less important
as the amount of data grows. Most of the improvement comes between the dataset
sizes of 1,000 and 32,000, with AUC improving from 80\% to 89.7\% for VI. The
very largest datasets improve AUC slightly further, with VI fitted to the entire
dataset reaching 90.4\% AUC. Maximum likelihood fitted to the entire dataset
separately for each species achieves a lower AUC of around 88.7\%. This gap
suggests that modelling the observation process with a hierarchical model is
able to improve predictions. \texttt{Stan} fitted to 1,000 checklists performs exactly as
well as MCMC, suggesting that \texttt{numpyro}'s estimates are very similar to
those made by \texttt{Stan}.

The middle panel, comparing log likelihood, shows a similar trend. There appears
to be a slight downward trend as the number of checklists grows beyond 32,000 in
VI, which may be because the number of fixed draws to approximate the objective
is reduced to allow the model to be loaded in memory. The difference is very
small, however, and unlikely to be important in practice. Maximum likelihood
performs worse on log likelihood than on AUC: even hierarchical models fitted to
a random sample of 2,000 checklists outperform maximum likelihood fitted to the
entire dataset, indicating that calibration is substantially improved using the
Bayesian multi-species models.

Finally, the bottom panel compares the runtime of the different approaches. The
slowest of all approaches is \texttt{Stan} fitted to the 1,000 checklists. We
note that while we attempted to write the \texttt{Stan} model efficiently, more
could be done to improve its performance. Our purpose here is not to provide a
benchmark comparing \texttt{Stan} and \texttt{numpyro}, but to use \texttt{Stan}
as a trusted method to ensure \texttt{numpyro} is providing reliable
estimates. \texttt{numpyro}'s runtime increased as the number of checklists
grows, and fitting 32,000 checklists took around 18 hours. Variational inference
was substantially faster, taking less than two hours to fit the largest
dataset. Given the similar AUC and log-likelihood metrics but dramatically
faster runtime, we believe VI to be the most compelling inference approach and
analyse the detectability estimates of the model fitted to the full dataset
using VI in the rest of the manuscript.

\begin{figure}
  \includegraphics[width=\textwidth]{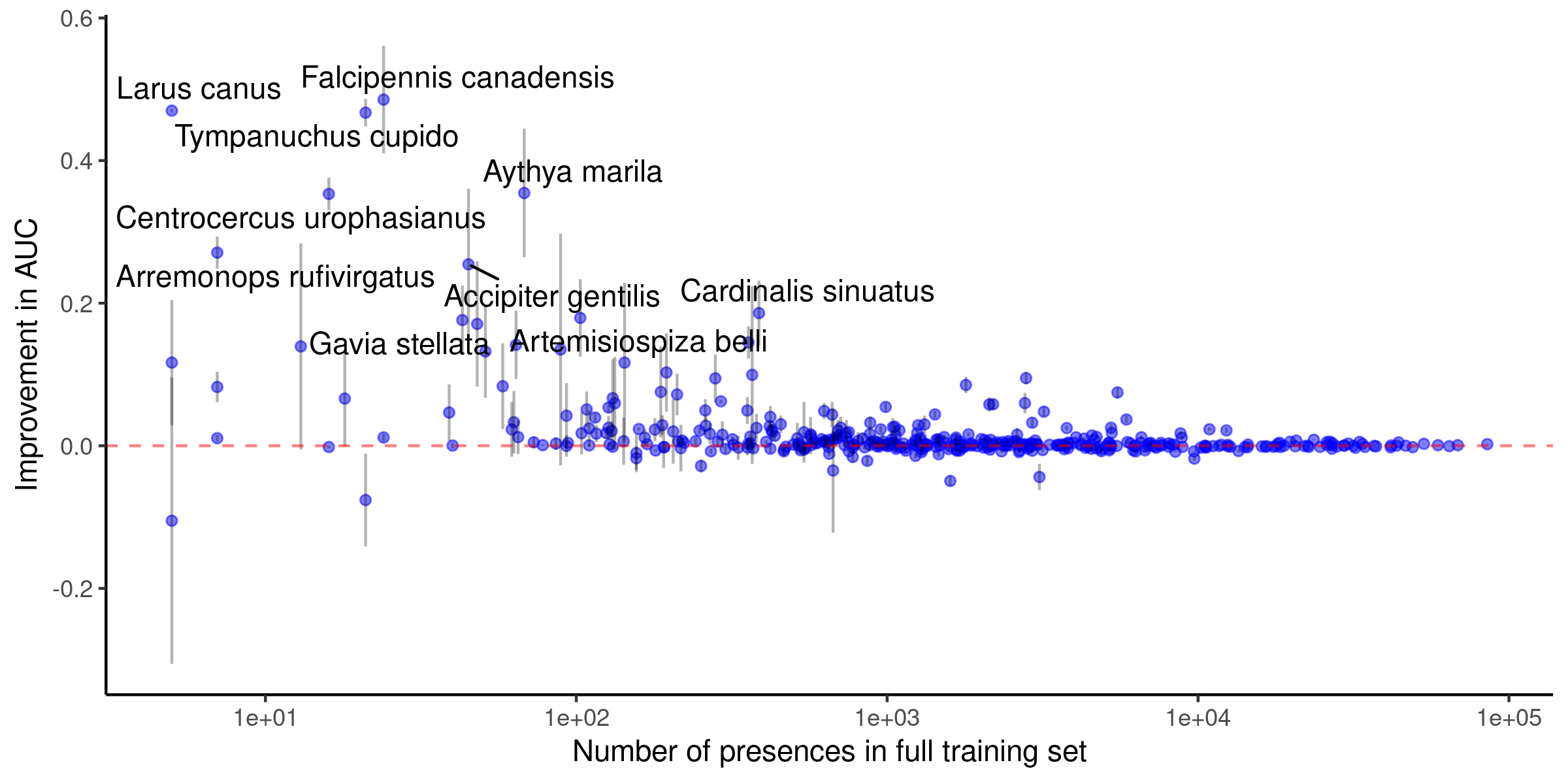}
  \caption{The figure shows the improvement in AUC for the multi-species
    occupancy detection model as compared to maximum likelihood fitted to each
    species separately. Both are fitted using the entire training set. Each
    point is a species in the dataset. Please note that the x-axis is shown on a
    logarithmic scale. The ten species with the largest absolute difference are
    annotated. Error bars show two standard errors estimated using
    bootstrapping.}
  \label{fig:auc-improvement}
\end{figure}

In the introduction, we suggested that multi-species occupancy detection models
may be particularly useful to predict rarely-observed species due to their
ability to borrow strength. To investigate this, we calculated the improvement
in AUC for each species when using the MSOD model fitted using VI compared to
maximum likelihood fitted separately to each
species. \autoref{fig:auc-improvement} shows these AUC improvements as a
function of the number of observations for each species in the training
dataset. The figure suggests that for the most common species with more than
1,000 presences or so, the single-species approach (using maximum likelihood)
and the multi-species approaches generally achieve a similar AUC. For rarer
species however, there appears to be a clear benefit to partially pooling the
observation process: almost all species show a positive improvement in AUC, and
some improve a considerable amount. Of those that show reduced AUC, almost all
error bars overlap with zero, indicating that the proposed model almost always
performs as well or better than maximum likelihood.

\subsection{Detectability estimates}

\begin{table}
  \centering
  \begin{tabular}{lrr}
\toprule
\textbf{Coefficient} &  \textbf{Group mean} &    \textbf{Group sd} \\
\midrule
Intercept                        & -1.83 &  0.86 \\
\midrule
Protocol: Stationary & -0.43 &  0.38 \\
Protocol: Traveling & -0.03 &  0.00 \\
\midrule
Daytime: Early morning    & -0.16 &  0.27 \\
Daytime: Late morning     & -0.32 &  0.27 \\
Daytime: Mid-day          & -0.46 &  0.33 \\
Daytime: Early evening    & -0.49 &  0.35 \\
Daytime: Late evening     & -0.49 &  0.35 \\
Daytime: Dusk             & -1.79 &  1.18 \\
Daytime: Night            & -1.78 &  1.03 \\
\midrule
Dominant Land Cover: Developed & -0.90 &  0.72 \\
Dominant Land Cover: Forest    & -0.44 &  0.86 \\
Dominant Land Cover: Water     & -0.47 &  0.60 \\
\midrule
Log duration                   &  0.45 &  0.20 \\
\bottomrule
\end{tabular}
\caption{Group means and standard deviations for the detectability model
  inferred by the variational inference model fitted to all training checklists.}
\label{tab:detectability-table}
\end{table}

\autoref{tab:detectability-table} shows the group means and standard deviations
inferred by the variational inference model fitted to all checklists in the
training set. The reference classes for the categorical variables are ``Area''
for protocol, ``dawn'' for daytimes, and ``Other'' for land cover.  The log
duration is standardised so that a log duration of zero corresponds to about 22
minutes spent birding. The group intercept of $-1.83 \pm 0.86$ suggests that a
typical bird species has a $\textrm{logit}^{-1}(-1.83) \approx 13.8\%$
probability of being detected when 22 minutes is spent following the Area
protocol, but that a probability of
$\textrm{logit}^{-1}(-1.83 - 2\times 0.86) \approx 2.8\%$ or as high as
$\textrm{logit}^{-1}(-1.83 + 2\times 0.86) \approx 47.3\%$ would not be very
surprising. In fact, the species with the lowest intercept is the Northern
goshawk (\emph{Accipiter gentilis}), whose intercept is
$\textrm{logit}^{-1}(-5.99) \approx 0.25\%$, indicating that this species is
very unlikely to be detected even if present. The American robin (\emph{Turdus
  migratorius}) has the highest intercept at
$\textrm{logit}^{-1}(0.77) \approx 68.4\%$. The group mean for log duration is
positive, indicating that in general, a longer time spent birding raises the
chance of detection, as one might expect.

The means of -0.43 and -0.03 imply that compared to the Area protocol, bird
species are somewhat less likely to be detected when following the Stationary
protocol, and roughly as likely when following the Traveling protocol. The group
standard deviation of 0.38 suggests that there is variability among the
Stationary coefficients by species, and indeed they range from -1.37 for 
Bell's vireo (\emph{Vireo bellii}) to 1.14 for the Ruby-throated
hummingbird (\emph{Archilochus colubris}). It is likely that the stationary
protocol is often followed by observers in their back yards, where hummingbirds
are often attracted to feeders, which may explain this high coefficient. By
contrast, the group standard deviation for the Traveling protocol is zero,
indicating that the variational inference procedure estimates no difference
among birds here compared to the Area protocol. 

\begin{figure}
  \includegraphics[width=\textwidth]{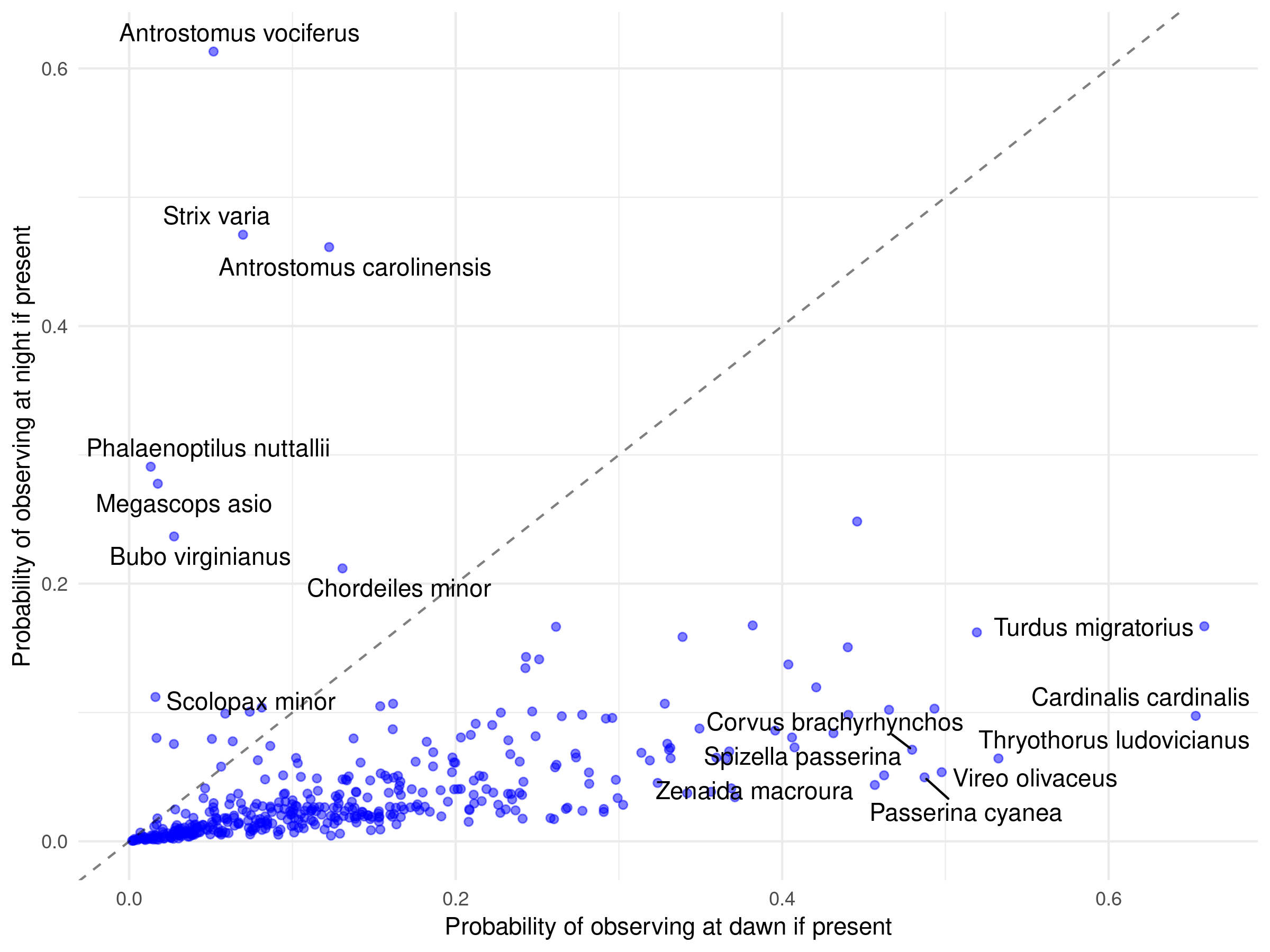}
  \caption{Estimated detection probabilities for bird species on a 22-minute
    forest walk at dawn (x-axis) compared to night (y-axis). Bird species on the
    dashed line would have an equal probability of being detected in both
    scenarios.}
  \label{fig:detection-probs}
\end{figure}

All daytime group means are smaller than zero, suggesting that most bird species
are most easily detected at dawn, the reference class, which is consistent with
common knowledge. Birds are typically considerably less likely to be detected at
dusk and night, but the group standard deviations of 1.18 and 1.03,
respectively, indicate that there is considerable variation by species.
\autoref{fig:detection-probs} investigates this in more depth. Species above the
dashed line are more likely to be detected at night, while those below it (the
majority) are more likely to be detected at dawn. For some species, the
difference is dramatic: the Eastern Whip-poor-will (\emph{Antrostomus
  vociferus}), for instance, is estimated to be very likely to be detected at
night (61.3\%) but quite unlikely to be detected at dawn (5.2\%). The other
species more likely to be detected at night are generally other noctural species
such as owls and other nightjars, which appears reasonable.

All three land cover categories are estimated to decrease the probability of
detecting most birds. The largest amount of variability is estimated to be among
the forest cover class, at 0.86. Abert's Towhee (\emph{Melozone aberti}) is
estimated to be much less likely to be detected in forests (-3.88), while the
Varied Thrush (\emph{Ixoreus naevius}) is much more likely to be detected there
(+1.88).

\subsection{Expert map evaluation}

\begin{figure}
  \centering
  \includegraphics[width=0.8\textwidth]{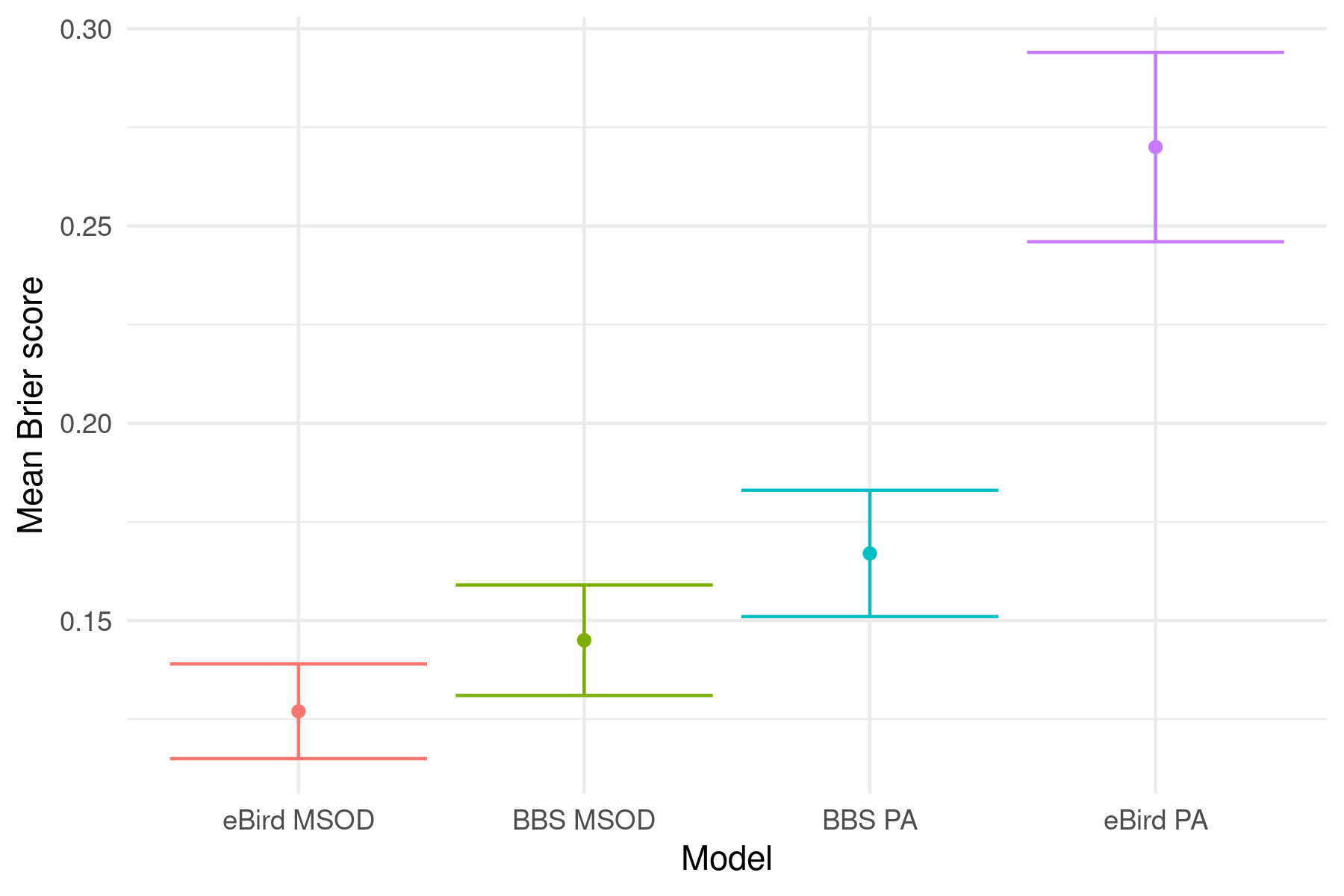}
  \caption{Comparison of mean Brier scores across species, treating the expert
    map as ground truth. A lower score indicates better agreement. The error
    bars indicate two standard errors.}
  \label{fig:mean-brier}
\end{figure}

\autoref{fig:mean-brier} shows the mean Brier score across species for each
combination of model (MSOD and PA) and dataset (eBird and BBS). A lower Brier
score indicates better agreement, on average, with the expert maps from
\cite{birdlife-maps}. The figure shows that there is a large improvement in
agreement when using MSOD models on eBird compared to PA, emphasising the
importance of accounting for the observation process. The MSOD model fitted to
eBird has the lowest Brier score on average, but there is overlap in the
standard errors with MSOD fitted to BBS. Comparing the Brier scores between
these approaches, MSOD fitted to eBird had a lower error for 238 of 357 birds,
or 66.7\%. A sign test suggested that this advantage is very unlikely to be due
to chance ($p = 2.9 \times 10^{-10}$). Figure S2 in the supplementary material
investigates the differences between these approaches further by comparing the
Brier scores for these two approaches. Most scores are similar, but MSOD on
eBird may agree slightly better when predicting water birds. Fitting the MSOD
model slightly improves agreement on BBS compared to PA, but the change is not
as dramatic as for the eBird dataset. We also provide an interactive web
application\footnote{Accessible here:
  \url{https://martiningram.shinyapps.io/range_map_comparisons/}} where range
maps predicted by the four approaches can be compared for species of
interest. Overall, the results suggest that an MSOD model fitted to eBird data
agrees with expert maps at least as well as the same model fitted to the
high-quality survey data.

\begin{figure}
\centering
\begin{subfigure}[t]{\textwidth}
  \includegraphics[width=\textwidth]{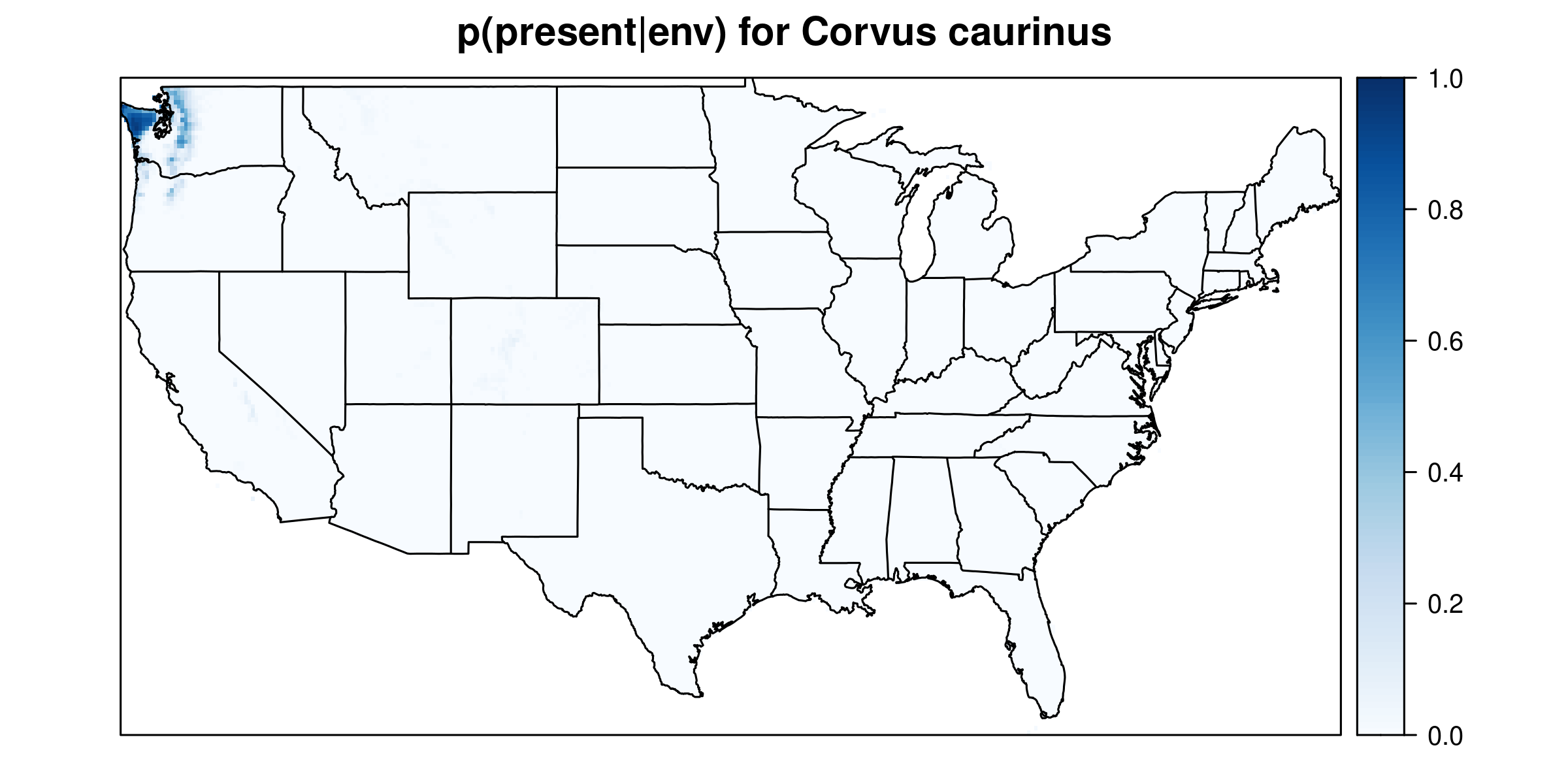}
  \caption{Probability of presence predicted by the proposed model.}
\end{subfigure}
\par\bigskip
\begin{subfigure}[t]{\textwidth}
   \includegraphics[width=1\linewidth]{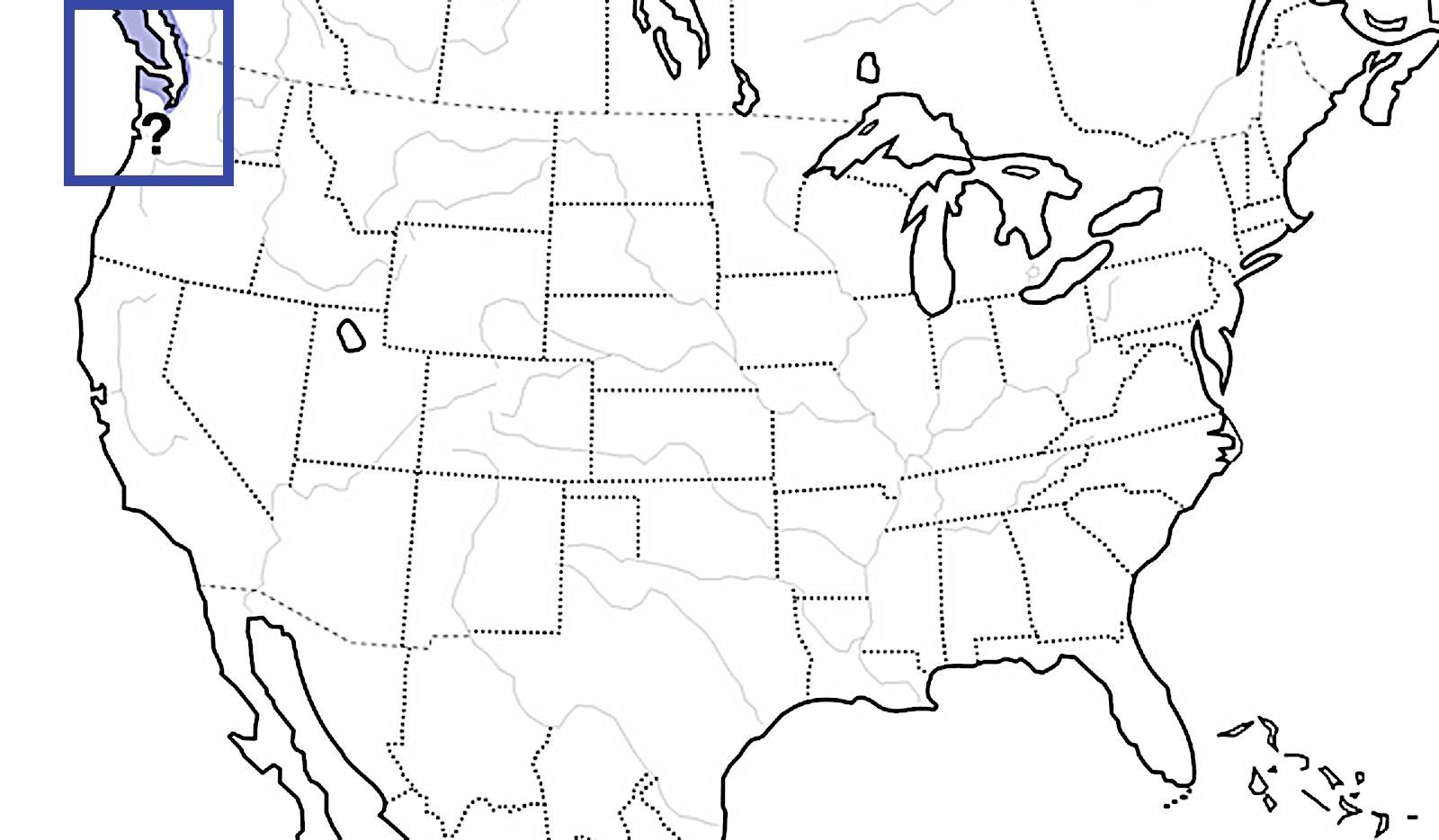}
   \caption{Expert map obtained from the
     \href{https://www.allaboutbirds.org/guide/Northwestern_Crow/maps-range}{All
       About Birds website}.}
\end{subfigure}
\caption{Predicted suitabilities and expert range map for
  \emph{Corvus caurinus}. Note the expert map has the species' distribution
  restricted to the North-West corner of the United States, and we have highlighted this with a blue box.}
\label{fig:comparison-crow}
\end{figure}

When comparing to BBS, it became apparent that the eBird data may be
particularly useful for predicting the distributions of rare species. In the
comparison above, we included only species that were observed at least eight
times in the BBS data, as species with fewer observations were unlikely to be
well estimated. Of the species fitted here, 47 were observed this rarely. About
80.8\% of these are observed at least 20 times in the eBird dataset, and 55.3\%
are observed more than 100 times, increasing the number of species for which
predictions could be made. \autoref{fig:comparison-crow} shows one such species:
the Northwestern Crow (\emph{Corvus caurinus}). This species had zero
observations in the BBS training set and a prediction was thus impossible. By
contrast, eBird contained 213 observations of \emph{Corvus caurinus} in the
training set, and the suitability map predicted in \autoref{fig:comparison-crow}
agrees well with the expert map.

We note that the predicted map in \autoref{fig:comparison-crow} shows the
posterior mean. The posterior mean is a reasonable point estimate as it
minimises the expected square error (see
e.g. \cite[p.161]{berger-decision-theory}), but it does not show the uncertainty
in the posterior. We discuss how to obtain and visualise the uncertainty in the
model's predictions in section H of the supplementary material.

\section{Discussion and conclusions}

The focus of this paper was to develop methods to scale multi-species occupancy
detection models to large datasets. We believe them to be a compelling model for
citizen science data, making the most of the wealth of data collected while
accounting for the heterogeneity in the observation process.

The ability to partially pool the coefficients in the observation process sets
this work apart from previous work applying occupancy detection models to
citizen science data. For example, \cite{johnston-best} fitted a single-species
occupancy detection model to a single bird species using eBird data. In this
work, we have shown that partial pooling can improve model predictions, both in
AUC and log likelihood.

Another approach taken in \cite{johnston-best} was to model the encounter rate
using random forests. Such models are able to model complicated environmental
responses, which may provide a better fit to the observed data. However, this
added flexibility comes at the cost of interpretability: the high-dimensional
interactions and non-linear functions fitted by random forests can be hard to
understand. Perhaps more importantly, unlike occupancy detection models, random
forests cannot estimate detection and occupancy probabilities separately,
further complicating interpretation. A more compelling alternative could be to
retain the occupancy detection modelling framework but use either generalised
additive \citep{hastie-tibshirani-gam} or Gaussian process models
\citep{mogp-paper} to allow for more complex environmental responses.

A key result of the paper is that, when evaluated against expert maps, the model
fitted to citizen science data performs at least as well as the same model
fitted to the high quality BBS dataset. As we have pointed out, compared to
survey data, citizen science data has the advantage of being more abundant,
allowing predictions to be made for rare species like the Northwestern crow
(\emph{Corvus caurinus}). In addition, while the BBS is only performed once a
year, eBird data is collected throughout, and the model presented could be
fitted to different time periods to investigate how distributions change over
the course of a year, or over time.

We see a number of ways to further refine the proposed model. The detection
probabilities could be made observer-specific, perhaps using an estimate of
observer skill such as that derived by \cite{accumulation-curves-ebird}. The
possibility of false positives could be taken into account, perhaps using the
methodology of \cite{ebird-common-misidentified} to identify species most at
risk of being confused. To estimate changes in occupancy over time, a dynamic
version of the proposed model could be developed \citep{dynamic-occ-det}. The
prior means could also be refined using species-specific information: for
example, a species indicator of whether it is nocturnal or not could be used to
set a prior expectation for the daytime coefficients.

As discussed previously, the model presented in this paper is non-spatial and treats presences in each cell as independent conditional on the environmental covariates. We have argued that this is a reasonable approximation, but we believe that extending the model to account for spatial autocorrelation would be worthwhile and should improve posterior variance estimates. A second useful spatial extension could be to allow the environmental coefficients to vary throughout each species' range \citep{non-stationary-sdm}. This could improve the model's predictions, as the study region is large, and species' environmental responses may differ across their range.

From the point of view of ecological research, we are excited about the possible applications of our approach. The ability to make estimates for species too rare in the BBS could be used to investigate their range shifts over time. To this end, the model could be fitted to data from previous years, and the range estimates could be compared. Similarly, because eBird is collected each month, changes in phenology, such as species migration patterns, could be assessed. While maps of species migration are already being produced with eBird data (see \cite{ebird-2009}, for example), our estimates corrected for observer bias may be preferable, especially when data are sparse. We also note that apart from the set of covariates chosen, our approach is general and not tied to the eBird dataset. It could thus be applied also to other citizen science datasets, and other species.

Finally, we would like to emphasise that the reason occupancy detection models can be
fitted to eBird data is the existence of the flag indicating that all species
that were observed were reported. Without this flag, eBird data would be
presence-only data, and at best we could only estimate relative suitability (see
\citet{fit-for-purpose}, for example). With the flag, as we have shown,
reasonable predictions can be made which we argued can even compete with those
based on systematic surveys. We believe that this has strong implications for
the design of citizen science protocols: citizen scientists should be asked to
report not only which species they observed, but also which species they were
looking for but did not observe.

\section*{Acknowledgements}

We would like to thank the Associate Editor and the two anonymous reviewers for their valuable feedback. We would also like to thank Alison Johnston for her helpful comments and suggestions on an early version of the manuscript.

This research was undertaken using the LIEF HPC-GPGPU Facility hosted at the
University of Melbourne. This Facility was established with the assistance of
LIEF Grant LE170100200. MI was partially supported by a Melbourne Research
Scholarship. NG was supported by an ARC DECRA fellowship (DE180100635).

\section*{Authors' contributions}

All authors conceived the ideas and methodology. MI implemented the analysis. MI
led writing the manuscript but all authors contributed significantly throughout
and gave final approval before submission.

\section*{Data accessibility}

\begin{itemize}
\item Scripts to fit the models discussed in this paper are available in a
  GitHub repository for python \url{https://github.com/martiningram/occu_py} and
  R (recommended) \url{https://github.com/martiningram/roccupy}.
\item The eBird dataset used in this paper is the eBird Basic Dataset (EBD),
  which is freely available at \url{https://ebird.org/data/download}
  \citep{ebird-2009}. The code we used to process this data is available at
  \url{https://github.com/martiningram/ebird-prep}.
\item The Breeding Bird Survey (BBS) dataset used in this paper is freely
  available from ScienceBase:
  \url{https://www.sciencebase.gov/catalog/item/5ea04e9a82cefae35a129d65}.
\end{itemize}

\bibliography{references}

\begin{thebibliography}{39}
\providecommand{\natexlab}[1]{#1}
\providecommand{\url}[1]{\texttt{#1}}
\expandafter\ifx\csname urlstyle\endcsname\relax
  \providecommand{\doi}[1]{doi: #1}\else
  \providecommand{\doi}{doi: \begingroup \urlstyle{rm}\Url}\fi

\bibitem[Altwegg and Nichols(2019)]{altwegg-2019}
Res Altwegg and James~D. Nichols.
\newblock Occupancy models for citizen-science data.
\newblock \emph{Methods in Ecology and Evolution}, 10\penalty0 (1):\penalty0
  8--21, 2019.
\newblock \doi{https://doi.org/10.1111/2041-210X.13090}.
\newblock URL
  \url{https://besjournals.onlinelibrary.wiley.com/doi/abs/10.1111/2041-210X.13090}.

\bibitem[Austin(2007)]{AUSTIN20071}
Mike Austin.
\newblock Species distribution models and ecological theory: A critical
  assessment and some possible new approaches.
\newblock \emph{Ecological Modelling}, 200\penalty0 (1):\penalty0 1 -- 19,
  2007.
\newblock ISSN 0304-3800.
\newblock \doi{https://doi.org/10.1016/j.ecolmodel.2006.07.005}.

\bibitem[Berger(1985)]{berger-decision-theory}
James~O Berger.
\newblock \emph{{Statistical decision theory and Bayesian analysis; 2nd ed.}}
\newblock Springer series in statistics. Springer, New York, 1985.
\newblock \doi{10.1007/978-1-4757-4286-2}.
\newblock URL \url{https://cds.cern.ch/record/1327974}.

\bibitem[{BirdLife International}(2020)]{birdlife-maps}
{BirdLife International}.
\newblock {Bird species distribution maps of the world. Version 2020.1.}
\newblock 2020.
\newblock URL \url{http://datazone.birdlife.org/species/requestdis}.

\bibitem[Bradbury et~al.(2018)Bradbury, Frostig, Hawkins, Johnson, Leary,
  Maclaurin, Necula, Paszke, Vander{P}las, Wanderman-{M}ilne, and
  Zhang]{jax2018github}
James Bradbury, Roy Frostig, Peter Hawkins, Matthew~James Johnson, Chris Leary,
  Dougal Maclaurin, George Necula, Adam Paszke, Jake Vander{P}las, Skye
  Wanderman-{M}ilne, and Qiao Zhang.
\newblock {JAX}: composable transformations of {P}ython+{N}um{P}y programs,
  2018.
\newblock URL \url{http://github.com/google/jax}.

\bibitem[Dickinson et~al.(2010)Dickinson, Zuckerberg, and
  Bonter]{citizen-science-research-tool}
Janis~L. Dickinson, Benjamin Zuckerberg, and David~N. Bonter.
\newblock Citizen science as an ecological research tool: Challenges and
  benefits.
\newblock \emph{Annual Review of Ecology, Evolution, and Systematics},
  41\penalty0 (1):\penalty0 149--172, 2010.
\newblock \doi{10.1146/annurev-ecolsys-102209-144636}.
\newblock URL \url{https://doi.org/10.1146/annurev-ecolsys-102209-144636}.

\bibitem[Fielding and Bell(1997)]{fielding1997review}
Alan~H Fielding and John~F Bell.
\newblock A review of methods for the assessment of prediction errors in
  conservation presence/absence models.
\newblock \emph{Environmental Conservation}, 24\penalty0 (1):\penalty0 38--49,
  1997.

\bibitem[Fiske and Chandler(2011)]{unmarked}
Ian Fiske and Richard Chandler.
\newblock {{unmarked}: An {R} Package for Fitting Hierarchical Models of
  Wildlife Occurrence and Abundance}.
\newblock \emph{Journal of Statistical Software}, 43\penalty0 (10):\penalty0
  1--23, 2011.
\newblock URL \url{http://www.jstatsoft.org/v43/i10/}.

\bibitem[Giordano et~al.(2018)Giordano, Broderick, and Jordan]{giordano-robust}
Ryan Giordano, Tamara Broderick, and Michael~I. Jordan.
\newblock {Covariances, Robustness, and Variational Bayes}.
\newblock \emph{Journal of Machine Learning Research}, 19\penalty0
  (51):\penalty0 1--49, 2018.
\newblock URL \url{http://jmlr.org/papers/v19/17-670.html}.

\bibitem[Gneiting and Raftery(2007)]{gneiting-scoring}
Tilmann Gneiting and Adrian~E Raftery.
\newblock Strictly {P}roper {S}coring {R}ules, {P}rediction, and {E}stimation.
\newblock \emph{Journal of the American Statistical Association}, 102\penalty0
  (477):\penalty0 359--378, 2007.
\newblock \doi{10.1198/016214506000001437}.
\newblock URL \url{https://doi.org/10.1198/016214506000001437}.

\bibitem[Golding(2019)]{Golding2019}
Nick Golding.
\newblock {greta: simple and scalable statistical modelling in R}.
\newblock \emph{Journal of Open Source Software}, 4\penalty0 (40):\penalty0
  1601, 2019.
\newblock \doi{10.21105/joss.01601}.
\newblock URL \url{https://doi.org/10.21105/joss.01601}.

\bibitem[Guillera-Arroita(2017)]{arroita-2017}
Gurutzeta Guillera-Arroita.
\newblock {Modelling of species distributions, range dynamics and communities
  under imperfect detection: advances, challenges and opportunities}.
\newblock \emph{Ecography}, 40\penalty0 (2):\penalty0 281--295, 2017.
\newblock \doi{https://doi.org/10.1111/ecog.02445}.
\newblock URL \url{https://onlinelibrary.wiley.com/doi/abs/10.1111/ecog.02445}.

\bibitem[Guillera-Arroita et~al.(2015)Guillera-Arroita, Lahoz-Monfort, Elith,
  Gordon, Kujala, Lentini, McCarthy, Tingley, and Wintle]{fit-for-purpose}
Gurutzeta Guillera-Arroita, José~J. Lahoz-Monfort, Jane Elith, Ascelin Gordon,
  Heini Kujala, Pia~E. Lentini, Michael~A. McCarthy, Reid Tingley, and
  Brendan~A. Wintle.
\newblock {Is my species distribution model fit for purpose? Matching data and
  models to applications}.
\newblock \emph{Global Ecology and Biogeography}, 24\penalty0 (3):\penalty0
  276--292, 2015.
\newblock \doi{https://doi.org/10.1111/geb.12268}.
\newblock URL \url{https://onlinelibrary.wiley.com/doi/abs/10.1111/geb.12268}.

\bibitem[Guillera-Arroita et~al.(2019)Guillera-Arroita, Kéry, and
  Lahoz-Monfort]{arroita-occ-det2019}
Gurutzeta Guillera-Arroita, Marc Kéry, and José~J. Lahoz-Monfort.
\newblock {Inferring species richness using multispecies occupancy modeling:
  Estimation performance and interpretation}.
\newblock \emph{Ecology and Evolution}, 9\penalty0 (2):\penalty0 780--792,
  2019.
\newblock \doi{https://doi.org/10.1002/ece3.4821}.
\newblock URL \url{https://onlinelibrary.wiley.com/doi/abs/10.1002/ece3.4821}.

\bibitem[Harris(2015)]{mistnet}
David~J. Harris.
\newblock Generating realistic assemblages with a joint species distribution
  model.
\newblock \emph{Methods in Ecology and Evolution}, 6\penalty0 (4):\penalty0
  465--473, 2015.
\newblock \doi{https://doi.org/10.1111/2041-210X.12332}.
\newblock URL
  \url{https://besjournals.onlinelibrary.wiley.com/doi/abs/10.1111/2041-210X.12332}.

\bibitem[Hastie and Tibshirani(1986)]{hastie-tibshirani-gam}
Trevor Hastie and Robert Tibshirani.
\newblock {Generalized Additive Models}.
\newblock \emph{Statistical Science}, 1\penalty0 (3):\penalty0 297 -- 310,
  1986.
\newblock \doi{10.1214/ss/1177013604}.
\newblock URL \url{https://doi.org/10.1214/ss/1177013604}.

\bibitem[Hijmans et~al.(2005)Hijmans, Cameron, Parra, Jones, and
  Jarvis]{hijmans2005very}
Robert~J. Hijmans, Susan~E. Cameron, Juan~L. Parra, Peter~G. Jones, and Andy
  Jarvis.
\newblock Very high resolution interpolated climate surfaces for global land
  areas.
\newblock \emph{International Journal of Climatology}, 25\penalty0
  (15):\penalty0 1965--1978, 2005.
\newblock \doi{10.1002/joc.1276}.

\bibitem[Hodges and Reich(2010)]{spatial-confounding}
James~S. Hodges and Brian~J. Reich.
\newblock Adding spatially-correlated errors can mess up the fixed effect you
  love.
\newblock \emph{The American Statistician}, 64\penalty0 (4):\penalty0 325--334,
  2010.
\newblock \doi{10.1198/tast.2010.10052}.
\newblock URL \url{https://doi.org/10.1198/tast.2010.10052}.

\bibitem[Hoffman and Gelman(2014)]{hoffman14a}
Matthew~D. Hoffman and Andrew Gelman.
\newblock {The No-U-Turn Sampler: Adaptively Setting Path Lengths in
  Hamiltonian Monte Carlo}.
\newblock \emph{Journal of Machine Learning Research}, 15\penalty0
  (47):\penalty0 1593--1623, 2014.
\newblock URL \url{http://jmlr.org/papers/v15/hoffman14a.html}.

\bibitem[Homer et~al.(2012)Homer, Fry, and Barnes]{homer2012national}
Collin~H Homer, Joyce~A Fry, and Christopher~A Barnes.
\newblock {The National Land Cover Database}.
\newblock \emph{US Geological Survey Fact Sheet}, 3020\penalty0 (4):\penalty0
  1--4, 2012.

\bibitem[Ingram et~al.(2020)Ingram, Vukcevic, and Golding]{mogp-paper}
Martin Ingram, Damjan Vukcevic, and Nick Golding.
\newblock {Multi-output Gaussian processes for species distribution modelling}.
\newblock \emph{Methods in Ecology and Evolution}, 11\penalty0 (12):\penalty0
  1587--1598, 2020.
\newblock \doi{https://doi.org/10.1111/2041-210X.13496}.
\newblock URL
  \url{https://besjournals.onlinelibrary.wiley.com/doi/abs/10.1111/2041-210X.13496}.

\bibitem[Johnson et~al.(2013)Johnson, Conn, Hooten, Ray, and
  Pond]{spatial-occ-det}
Devin~S. Johnson, Paul~B. Conn, Mevin~B. Hooten, Justina~C. Ray, and Bruce~A.
  Pond.
\newblock Spatial occupancy models for large data sets.
\newblock \emph{Ecology}, 94\penalty0 (4):\penalty0 801--808, 2013.
\newblock \doi{https://doi.org/10.1890/12-0564.1}.
\newblock URL
  \url{https://esajournals.onlinelibrary.wiley.com/doi/abs/10.1890/12-0564.1}.

\bibitem[Johnston et~al.(2019)Johnston, Hochachka, Strimas-Mackey, Gutierrez,
  Robinson, Miller, Auer, Kelling, and Fink]{johnston-best}
A~Johnston, WM~Hochachka, ME~Strimas-Mackey, V~Ruiz Gutierrez, OJ~Robinson,
  ET~Miller, T~Auer, ST~Kelling, and D~Fink.
\newblock {Best practices for making reliable inferences from citizen science
  data: case study using eBird to estimate species distributions}.
\newblock \emph{bioRxiv}, 2019.
\newblock \doi{10.1101/574392}.
\newblock URL \url{https://www.biorxiv.org/content/early/2019/03/13/574392}.

\bibitem[Kelling et~al.(2015)Kelling, Johnston, Hochachka, Iliff, Fink,
  Gerbracht, Lagoze, La~Sorte, Moore, Wiggins, Wong, Wood, and
  Yu]{accumulation-curves-ebird}
Steve Kelling, Alison Johnston, Wesley~M. Hochachka, Marshall Iliff, Daniel
  Fink, Jeff Gerbracht, Carl Lagoze, Frank~A. La~Sorte, Travis Moore, Andrea
  Wiggins, Weng-Keen Wong, Chris Wood, and Jun Yu.
\newblock {Can Observation Skills of Citizen Scientists Be Estimated Using
  Species Accumulation Curves?}
\newblock \emph{PLOS ONE}, 10\penalty0 (10):\penalty0 1--20, 10 2015.
\newblock \doi{10.1371/journal.pone.0139600}.
\newblock URL \url{https://doi.org/10.1371/journal.pone.0139600}.

\bibitem[Kery and Royle(2009)]{Kery2009}
M.~Kery and J.~Andrew Royle.
\newblock \emph{{Inference about species richness and community structure using
  species-specific occupancy models in the National Swiss Breeding Bird Survey
  MUB}}, pages 639--656.
\newblock Modeling demographic processes in marked populations. Springer, New
  York and London, 2009.
\newblock URL \url{http://pubs.er.usgs.gov/publication/5211455}.

\bibitem[Kucukelbir et~al.(2017)Kucukelbir, Tran, Ranganath, Gelman, and
  Blei]{advi-paper}
Alp Kucukelbir, Dustin Tran, Rajesh Ranganath, Andrew Gelman, and David~M.
  Blei.
\newblock {Automatic Differentiation Variational Inference}.
\newblock \emph{Journal of Machine Learning Research}, 18\penalty0
  (14):\penalty0 1--45, 2017.
\newblock URL \url{http://jmlr.org/papers/v18/16-107.html}.

\bibitem[Lawson et~al.(2014)Lawson, Hodgson, Wilson, and
  Richards]{lawson2014prevalence}
Callum~R Lawson, Jenny~A Hodgson, Robert~J Wilson, and Shane~A Richards.
\newblock {Prevalence, thresholds and the performance of presence--absence
  models}.
\newblock \emph{Methods in Ecology and Evolution}, 5\penalty0 (1):\penalty0
  54--64, 2014.

\bibitem[Lunn et~al.(2000)Lunn, Thomas, Best, and Spiegelhalter]{Lunn2000}
David~J. Lunn, Andrew Thomas, Nicky Best, and David Spiegelhalter.
\newblock {WinBUGS - A Bayesian modelling framework: Concepts, structure, and
  extensibility}.
\newblock \emph{Statistics and Computing}, 10\penalty0 (4):\penalty0 325--337,
  Oct 2000.
\newblock ISSN 1573-1375.
\newblock \doi{10.1023/A:1008929526011}.
\newblock URL \url{https://doi.org/10.1023/A:1008929526011}.

\bibitem[MacKenzie et~al.(2002)MacKenzie, Nichols, Lachman, Droege,
  Andrew~Royle, and Langtimm]{mackenzie-occ-det-2002}
Darryl~I. MacKenzie, James~D. Nichols, Gideon~B. Lachman, Sam Droege,
  J.~Andrew~Royle, and Catherine~A. Langtimm.
\newblock Estimating site occupancy rates when detection probabilities are less
  than one.
\newblock \emph{Ecology}, 83\penalty0 (8):\penalty0 2248--2255, 2002.
\newblock
  \doi{https://doi.org/10.1890/0012-9658(2002)083[2248:ESORWD]2.0.CO;2}.
\newblock URL
  \url{https://esajournals.onlinelibrary.wiley.com/doi/abs/10.1890/0012-9658%282002%29083%5B2248%3AESORWD%5D2.0.CO%3B2}.

\bibitem[Monnahan et~al.(2017)Monnahan, Thorson, and Branch]{faster-bayes-2016}
Cole~C. Monnahan, James~T. Thorson, and Trevor~A. Branch.
\newblock {Faster estimation of Bayesian models in ecology using Hamiltonian
  Monte Carlo}.
\newblock \emph{Methods in Ecology and Evolution}, 8\penalty0 (3):\penalty0
  339--348, 2017.
\newblock \doi{https://doi.org/10.1111/2041-210X.12681}.
\newblock URL
  \url{https://besjournals.onlinelibrary.wiley.com/doi/abs/10.1111/2041-210X.12681}.

\bibitem[Osborne et~al.(2007)Osborne, Foody, and
  Suárez-Seoane]{non-stationary-sdm}
Patrick~E. Osborne, Giles~M. Foody, and Susana Suárez-Seoane.
\newblock Non-stationarity and local approaches to modelling the distributions
  of wildlife.
\newblock \emph{Diversity and Distributions}, 13\penalty0 (3):\penalty0
  313--323, 2007.
\newblock ISSN 13669516, 14724642.
\newblock URL \url{http://www.jstor.org/stable/4539924}.

\bibitem[Pardieck et~al.(2020)Pardieck, Ziolkowski~Jr, Lutmerding, Aponte, and
  Hudson]{pardieck2020north}
K.L. Pardieck, D.J. Ziolkowski~Jr, M.~Lutmerding, V.I. Aponte, and M-A.R.
  Hudson.
\newblock {North {A}merican {B}reeding {B}ird {S}urvey {D}ataset 1966-2019:
  U.S. Geological Survey data release}.
\newblock \emph{US Geological Survey, Patuxent Wildlife Research Center}, 2020.
\newblock \doi{10.5066/P9J6QUF6}.

\bibitem[Phan et~al.(2019)Phan, Pradhan, and Jankowiak]{phan2019composable}
Du~Phan, Neeraj Pradhan, and Martin Jankowiak.
\newblock {Composable Effects for Flexible and Accelerated Probabilistic
  Programming in NumPyro}.
\newblock \emph{arXiv preprint arXiv:1912.11554}, 2019.

\bibitem[Royle and Kéry(2007)]{dynamic-occ-det}
J.~Andrew Royle and Marc Kéry.
\newblock {A Bayesian state-space formulation of dynamic occupancy models}.
\newblock \emph{Ecology}, 88\penalty0 (7):\penalty0 1813--1823, 2007.
\newblock \doi{https://doi.org/10.1890/06-0669.1}.
\newblock URL
  \url{https://esajournals.onlinelibrary.wiley.com/doi/abs/10.1890/06-0669.1}.

\bibitem[Royle et~al.(2007)Royle, Dorazio, and Link]{royle-multinomial-2007}
J.~Andrew Royle, Robert~M Dorazio, and William~A Link.
\newblock {Analysis of Multinomial Models With Unknown Index Using Data
  Augmentation}.
\newblock \emph{Journal of Computational and Graphical Statistics}, 16\penalty0
  (1):\penalty0 67--85, 2007.
\newblock \doi{10.1198/106186007X181425}.
\newblock URL \url{https://doi.org/10.1198/106186007X181425}.

\bibitem[{Stan Development Team}(2020)]{rstan}
{Stan Development Team}.
\newblock {RStan}: the {R} interface to {Stan}, 2020.
\newblock URL \url{http://mc-stan.org/}.
\newblock R package version 2.21.2.

\bibitem[Sullivan et~al.(2009)Sullivan, Wood, Iliff, Bonney, Fink, and
  Kelling]{ebird-2009}
Brian~L. Sullivan, Christopher~L. Wood, Marshall~J. Iliff, Rick~E. Bonney,
  Daniel Fink, and Steve Kelling.
\newblock {eBird: A citizen-based bird observation network in the biological
  sciences}.
\newblock \emph{Biological Conservation}, 142\penalty0 (10):\penalty0
  2282--2292, 2009.
\newblock ISSN 0006-3207.
\newblock \doi{https://doi.org/10.1016/j.biocon.2009.05.006}.
\newblock URL
  \url{https://www.sciencedirect.com/science/article/pii/S000632070900216X}.

\bibitem[Valavi et~al.(2019)Valavi, Elith, Lahoz-Monfort, and
  Guillera-Arroita]{valavi2019block}
Roozbeh Valavi, Jane Elith, Jos{\'e}~J Lahoz-Monfort, and Gurutzeta
  Guillera-Arroita.
\newblock block{C}{V}: An {R} package for generating spatially or
  environmentally separated folds for k-fold cross-validation of species
  distribution models.
\newblock \emph{Methods in Ecology and Evolution}, 10\penalty0 (2):\penalty0
  225--232, 2019.

\bibitem[Yu et~al.(2014)Yu, Hutchinson, and Wong]{ebird-common-misidentified}
Jun Yu, Rebecca Hutchinson, and Weng-Keen Wong.
\newblock {A Latent Variable Model for Discovering Bird Species Commonly
  Misidentified by Citizen Scientists}.
\newblock \emph{Proceedings of the AAAI Conference on Artificial Intelligence},
  28\penalty0 (1), Jun. 2014.
\newblock URL \url{https://ojs.aaai.org/index.php/AAAI/article/view/8763}.

\end{thebibliography}


\begin{thebibliography}{5}
\providecommand{\natexlab}[1]{#1}
\providecommand{\url}[1]{\texttt{#1}}
\expandafter\ifx\csname urlstyle\endcsname\relax
  \providecommand{\doi}[1]{doi: #1}\else
  \providecommand{\doi}{doi: \begingroup \urlstyle{rm}\Url}\fi

\bibitem[Bradbury et~al.(2018)Bradbury, Frostig, Hawkins, Johnson, Leary,
  Maclaurin, Necula, Paszke, Vander{P}las, Wanderman-{M}ilne, and
  Zhang]{jax2018github}
James Bradbury, Roy Frostig, Peter Hawkins, Matthew~James Johnson, Chris Leary,
  Dougal Maclaurin, George Necula, Adam Paszke, Jake Vander{P}las, Skye
  Wanderman-{M}ilne, and Qiao Zhang.
\newblock {JAX}: composable transformations of {P}ython+{N}um{P}y programs,
  2018.
\newblock URL \url{http://github.com/google/jax}.

\bibitem[Fiske and Chandler(2011)]{unmarked}
Ian Fiske and Richard Chandler.
\newblock {{unmarked}: An {R} Package for Fitting Hierarchical Models of
  Wildlife Occurrence and Abundance}.
\newblock \emph{Journal of Statistical Software}, 43\penalty0 (10):\penalty0
  1--23, 2011.
\newblock URL \url{http://www.jstatsoft.org/v43/i10/}.

\bibitem[Giordano et~al.(2018)Giordano, Broderick, and Jordan]{giordano-robust}
Ryan Giordano, Tamara Broderick, and Michael~I. Jordan.
\newblock {Covariances, Robustness, and Variational Bayes}.
\newblock \emph{Journal of Machine Learning Research}, 19\penalty0
  (51):\penalty0 1--49, 2018.
\newblock URL \url{http://jmlr.org/papers/v19/17-670.html}.

\bibitem[Kucukelbir et~al.(2017)Kucukelbir, Tran, Ranganath, Gelman, and
  Blei]{advi-paper}
Alp Kucukelbir, Dustin Tran, Rajesh Ranganath, Andrew Gelman, and David~M.
  Blei.
\newblock {Automatic Differentiation Variational Inference}.
\newblock \emph{Journal of Machine Learning Research}, 18\penalty0
  (14):\penalty0 1--45, 2017.
\newblock URL \url{http://jmlr.org/papers/v18/16-107.html}.

\bibitem[Virtanen et~al.(2020)Virtanen, Gommers, Oliphant, Haberland, Reddy,
  Cournapeau, Burovski, Peterson, Weckesser, Bright, {van der Walt}, Brett,
  Wilson, Millman, Mayorov, Nelson, Jones, Kern, Larson, Carey, Polat, Feng,
  Moore, {VanderPlas}, Laxalde, Perktold, Cimrman, Henriksen, Quintero, Harris,
  Archibald, Ribeiro, Pedregosa, {van Mulbregt}, and {SciPy 1.0
  Contributors}]{2020SciPy-NMeth}
Pauli Virtanen, Ralf Gommers, Travis~E. Oliphant, Matt Haberland, Tyler Reddy,
  David Cournapeau, Evgeni Burovski, Pearu Peterson, Warren Weckesser, Jonathan
  Bright, St{\'e}fan~J. {van der Walt}, Matthew Brett, Joshua Wilson, K.~Jarrod
  Millman, Nikolay Mayorov, Andrew R.~J. Nelson, Eric Jones, Robert Kern, Eric
  Larson, C~J Carey, {\.I}lhan Polat, Yu~Feng, Eric~W. Moore, Jake
  {VanderPlas}, Denis Laxalde, Josef Perktold, Robert Cimrman, Ian Henriksen,
  E.~A. Quintero, Charles~R. Harris, Anne~M. Archibald, Ant{\^o}nio~H. Ribeiro,
  Fabian Pedregosa, Paul {van Mulbregt}, and {SciPy 1.0 Contributors}.
\newblock {{{SciPy} 1.0: Fundamental Algorithms for Scientific Computing in
  Python}}.
\newblock \emph{Nature Methods}, 17:\penalty0 261--272, 2020.
\newblock \doi{10.1038/s41592-019-0686-2}.

\end{thebibliography}
\bibliographystyle{plainnat}

\end{document}


\maketitle

\begin{appendices}

\renewcommand{\thefigure}{S\arabic{figure}}
\setcounter{figure}{0}

\section{Derivation of observed-data likelihood}

The likelihood factorises across sites and species. Focusing on just one site
and species:
\begin{align}
  p(y_{ij}, s_{ij1}, ..., s_{ijK_i} \mid \theta) = p(y_{ij} \mid x_i) \prod_{k=1}^{K_i} p(s_{ijk} \mid y_{ij}, x_{ijk}^{\text{(obs)}}).
\end{align}

Since $y_{ij}$ is not observed, we sum it out to get the observed-data likelihood:
\begin{align}
  p(s_{ij1}, ..., s_{ijK_i} \mid \theta) =
  &p(y_{ij} = 0 \mid x_i) \prod_{k=1}^{K_i} p(s_{ijk} \mid y_{ij} = 0,
    x_{ik}^{\text{(obs)}}) +  \\
  &p(y_{ij} = 1 \mid x_i) \prod_{k=1}^{K_i} p(s_{ijk} \mid y_{ij} =
    1, x_{ik}^{\text{(obs)}})
\end{align}

We now focus on each term in turn. Firstly:
\begin{align}
p(y_{ij} = 0 \mid x_i) \prod_{k=1}^{K_i} p(s_{ijk} \mid y_{ij} = 0, x_{ik}^{\text{(obs)}}) =
  (1 - \Psi_{ij}) \prod_{k=1}^{K_i} p(s_{ijk} \mid y_{ij} = 0, x_{ik}^{\text{(obs)}}) 
\label{eq:y-zero-term}
\end{align}

The first term, $1 - \Psi_{ij}$, is given by the linear model with a logit link
described in the main text. The second term is a product. Focusing on one
checklist $k$, it is:
\begin{align}
p(s_{ijk} \mid y_{ij} = 0, x_{ik}^{\text{(obs)}}).
\end{align}

This is the probability of making the observation $s_{ijk}$ when the species is
truly absent. A common assumption, and one made here, is that there are no false
positives. Hence:
\begin{align}
p(s_{ijk} = 1 \mid y_{ij} = 0, x_{ik}^{\text{(obs)}}) = 0.
\end{align}

This implies that \autoref{eq:y-zero-term} is non-zero only when the
species was not observed on any of the $K_i$ checklists at site $i$. This is
intuitively reasonable: if one or more sightings of a species have been made at
a site and false positives are ruled out, we can be sure that the species is
present, meaning $y_{ij} = 0$ is impossible. This term can thus be written as:
\begin{align}
p(y_{ij} = 0 \mid x_i) \prod_{k=1}^{K_i} (1 - s_{ijk}),
\end{align}

which is equal to $p(y_{ij} = 0 \mid x_i)$ if the species was not observed on
any of the checklists, and $0$ otherwise.

The second term is:
\begin{align}
p(y_{ij} = 1 \mid x_i) \prod_{k=1}^{K_i} p(s_{ijk} \mid y_{ij} = 1, x_{ik}^{\text{(obs)}}) =
  \Psi_{ij} \prod_{k=1}^{K_i} (p_{ijk})^{s_{ijk}} (1 - p_{ijk})^{1 - s_{ijk}}.
\end{align}

Putting the two together and computing the product across sites and species
yields the final expression:
\begin{align}
  p(s \mid \theta) = \prod_{i=1}^{N} \prod_{j=1}^{J} \left[(1 - \Psi_{ij}) \prod_{k=1}^{K_i} (1 - s_{ijk}) +
  \Psi_{ij} \prod_{k=1}^{K_i} (p_{ijk})^{s_{ijk}} (1 - p_{ijk})^{1 - s_{ijk}} \right].
\end{align}

In practice, when implemented in software, this expression is computed on the log
scale to prevent underflow.

\section{Speed comparison with unmarked package}

To demonstrate the benefits of our proposed data structure, and of
gradient-based optimisation for maximum likelihood, we compare the runtime of
our proposed approach against that of the R package \texttt{unmarked}
\citep{unmarked}. We show in a separate section of the supplementary material
that the two approaches estimate the same coefficients; here, our focus is on
comparing the speed of the two implementations.

To perform this comparison, we select a single species, \emph{Cardinalis
  cardinalis}, and investigate how the model fitting time changes as the number
of checklists used is increased. The model used is slightly simpler than the one
discussed in the manuscript. It uses the worldclim variables with quadratic
terms, but no land cover, as the suitability covariates, and protocol type, time
of day, and log duration to model the detection probabilities.

\begin{figure}
  \includegraphics[width=\textwidth]{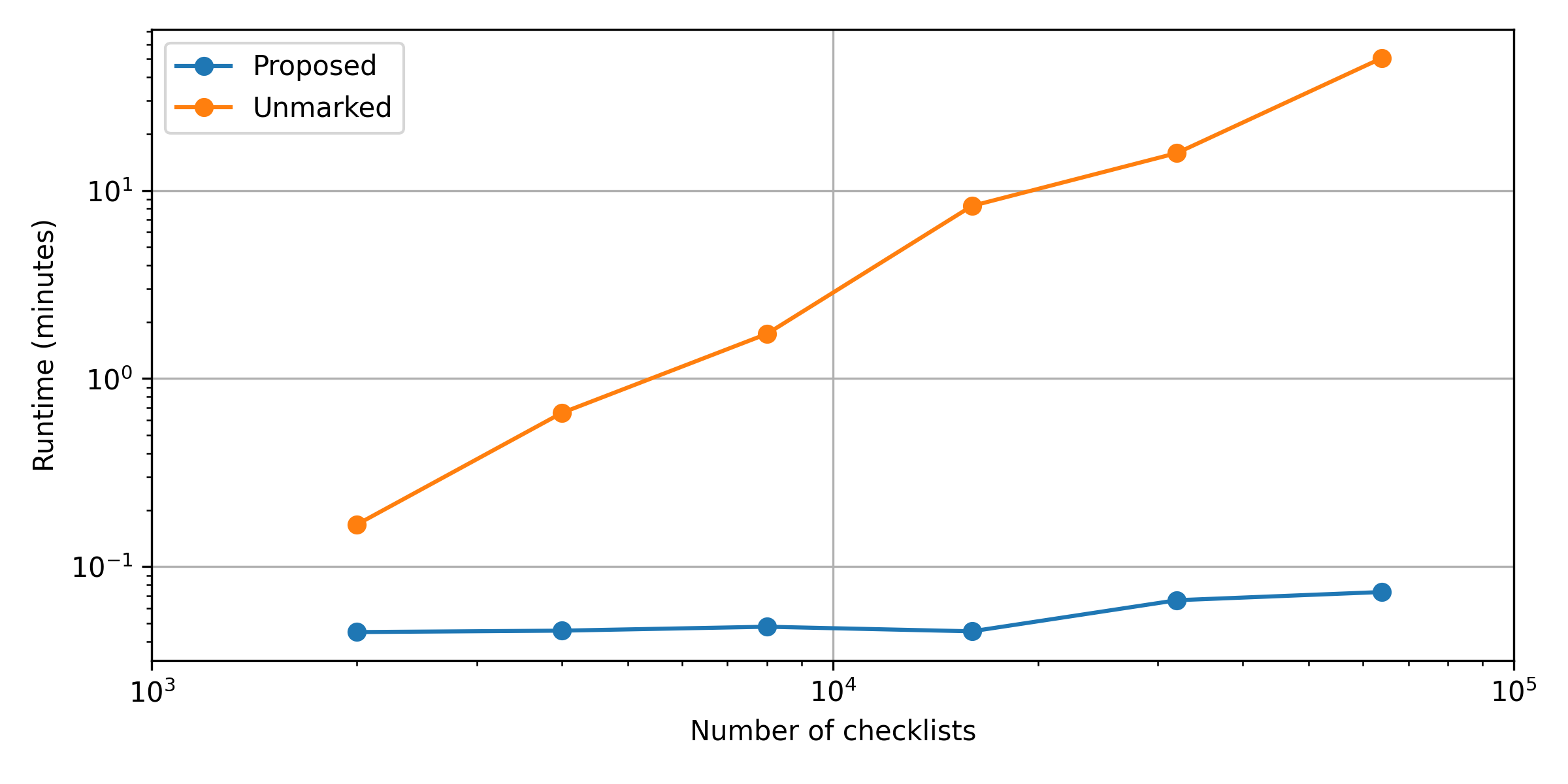}
  \caption{Runtime of the proposed maximum likelihood implementation compared to
    that in the R package \texttt{unmarked}.}
  \label{fig:runtime-unmarked-proposed}
\end{figure}

The result of the comparison is shown in
\autoref{fig:runtime-unmarked-proposed}. The speed increase when using the
proposed method is dramatic: the largest number of checklists considered here,
64,000, took about 38 minutes to fit using unmarked, but only about five seconds
using the proposed approach. As mentioned in the main text, this is due to a
combination of using a more efficient data structure, the availability of
gradients in the optimisation, and the use of a GPU. 

\section{Environmental covariate summaries}

\autoref{tab:env-cov-summaries} shows summaries of the environmental
covariates used.

\begin{table}
\centering
\begin{tabular}{lrr}
\toprule
\textbf{Covariate} &     \textbf{Mean} &      \textbf{Standard deviation} \\
\midrule
Annual Mean Temperature             &   105.84 &    51.73 \\
Mean Diurnal Range                  &   132.41 &    20.99 \\
Isothermality                       &    35.98 &     6.93 \\
Temperature Seasonality             &  8361.72 &  1875.86 \\
Max Temperature of Warmest Month    &   295.46 &    36.81 \\
Min Temperature of Coldest Month    &   -74.96 &    71.91 \\
Temperature Annual Range            &   370.42 &    60.24 \\
Mean Temperature of Wettest Quarter &   150.95 &    83.59 \\
Mean Temperature of Driest Quarter  &    56.87 &   112.97 \\
Mean Temperature of Warmest Quarter &   211.10 &    42.37 \\
Annual Precipitation                &   849.06 &   409.48 \\
Precipitation of Wettest Month      &   109.04 &    53.19 \\
Precipitation of Driest Month       &    38.22 &    27.87 \\
Precipitation Seasonality           &    36.07 &    20.88 \\
Precipitation of Warmest Quarter    &   230.97 &   118.17 \\
Precipitation of Coldest Quarter    &   193.70 &   161.21 \\
has\_open\_water                      &     0.08 &     0.27 \\
has\_deciduous\_forest                &     0.23 &     0.42 \\
has\_evergreen\_forest                &     0.21 &     0.41 \\
has\_mixed\_forest                    &     0.10 &     0.30 \\
has\_shrub\_or\_scrub                  &     0.21 &     0.41 \\
has\_grassland\_or\_herbaceous         &     0.16 &     0.36 \\
has\_pasture\_or\_hay                  &     0.14 &     0.35 \\
has\_cultivated\_crops                &     0.26 &     0.44 \\
has\_other                           &     0.02 &     0.13 \\
has\_developed                       &     0.10 &     0.30 \\
has\_wetlands                        &     0.12 &     0.32 \\
\bottomrule
\end{tabular}
\caption{Covariates used as part of the suitability model, together with their
  means and standard deviations across the entire dataset.}
\label{tab:env-cov-summaries}
\end{table}

\section{Automatic differentiation variational inference}

Here, we give a brief overview of the variational inference approach used in
this paper. For a more detailed description, we refer the reader to
\cite{advi-paper} and \cite{giordano-robust}.

In Bayesian inference, the goal is to estimate the posterior distribution of the
model parameters $\theta$ given the observed data $y$, $p(\theta \mid y)$. By
Bayes' rule, this is given by:
\begin{align*}
  p(\theta \mid y) = \frac{p(\theta) p(y \mid \theta)}{p(y)}.
\end{align*}
This posterior distribution is rarely available in closed form. Markov chain
Monte Carlo methods like Hamiltonian Monte Carlo construct a Markov chain whose
stationary distribution is $p(\theta \mid y)$ to sample from the posterior
distribution. Variational inference, on the other hand, approximates the
posterior distribution $p(\theta \mid y)$ with a distribution $q(\theta; \eta)$,
where $\eta$ are the parameters of this distribution. Here, we use independent
Gaussian distributions to approximate the posterior:
\begin{align*}
  q(\theta; \eta) = \prod_{i=1}^N q(\theta_i \mid \mu_i, \sigma_i^2).
\end{align*}
This means that the variational parameters $\eta$ are the collection of means
$\mu_i$ and variances $\sigma_i^2$. The goal is to find parameters $\eta^*$
which make $q(\theta; \eta)$ match the true posterior $p(\theta \mid y)$ as
closely as possible.

The way in which variational inference achieves this is by minimizing the
Kullback-Leibler (KL) divergence between $q(\theta; \eta)$ and $p(\theta \mid
y)$. The KL divergence can be written as:
\begin{align*}
  \textrm{KL}[q(\theta; \eta) \mid\mid p(\theta \mid y)] =
  - \mathbb{E}_{\theta \sim q(\theta; \eta)}[\log p(\theta) + \log p(y \mid \theta)]
  - \mathbb{H}[q(\theta; \eta)] + \textrm{constant},
\end{align*}
where the first term is the negative expected value of the unnormalised log
posterior distribution taken over the variational distribution, and the second
is the negative entropy of the variational approximation.

For factorising normal distributions, the second term is available in closed
form, leaving the first term. It can be rewritten as
\begin{align*}
  -\mathbb{E}_{z \sim \mathcal{N}(0, I)}[\log p(\theta(z, \eta)) + \log p(y \mid
  \theta(z, \eta)],
\end{align*}
where
\begin{align*}
  \theta(z, \eta) = \sigma \odot z + \mu,
\end{align*}
and $\odot$ denotes element-wise multiplication. This reparameterisation means
that the expectation does not depend on the parameters $\eta$, and so the
gradient of the expectation is equal to
\begin{align}
  -\mathbb{E}_{z \sim \mathcal{N}(0, I)}[\nabla (\log p(\theta(z, \eta)) + \log p(y \mid
  \theta(z, \eta))].
  \label{eq:expected-posterior}
\end{align}
This gradient is calculated using automatic differentiation, forming the ``AD''
in the abbreviation ``ADVI''.

In the original paper \citep{advi-paper}, this objective is optimised using
stochastic optimisation. Stochastic optimisation relies on first-order
optimisation, which can be challenging since it involves a number of tuning
parameters. For this reason, we use the variant proposed in
\cite{giordano-robust}, which fixes a set of $M$ draws for $z$ in
\autoref{eq:expected-posterior}. This incurs a small amount of bias, but allows
the use of second order methods to optimise the objective, which eliminate
tuning parameters in the optimisation. Choosing $M$ to be large reduces bias,
but also requires more memory. We implemented the code in the \texttt{JAX}
framework \citep{jax2018github} and optimise the objective using the Newton
conjugate gradient trust-region algorithm implemented in python's \texttt{scipy}
package \citep{2020SciPy-NMeth}.

\section{Details of observation covariates}

The daytime classifications, together with the percentage of checklists
classified as each, are the following:

\begin{itemize}
\item Dawn (2.6\%): One hour before sunrise
\item Dusk (1.1\%): One hour after sunset
\item Early morning (30.7\%): First three hours after sunrise
\item Late evening (11.5\%): Last three hours before sunset
\item Night (1.1\%): Time between dusk and dawn
\item Early evening (12.8\%): Three hours before late evening
\item Late morning (25.4\%): Three hours after early morning
\item Remaining times (14.8\%): Mid-day
\end{itemize}

\autoref{tab:land-cover-obs-fracs} shows the percentage of checklists
assigned to each land cover type.

\begin{table}
  \centering
\begin{tabular}{lr}
\toprule
{} &  \textbf{Percentage of checklists} \\
\midrule
Other  &                0.334 \\
Forest    &                0.299 \\
Developed &                0.280 \\
Water     &                0.087 \\
\bottomrule
\end{tabular}
\caption{The table shows the percentage of checklists classified as each land
  cover across the entire dataset.}
\label{tab:land-cover-obs-fracs}
\end{table}

The mean log duration across the dataset is 3.10, or about 22 minutes, with a
standard deviation of 1.29.

\section{Brier score comparison}

\begin{figure}
  \includegraphics[width=\textwidth]{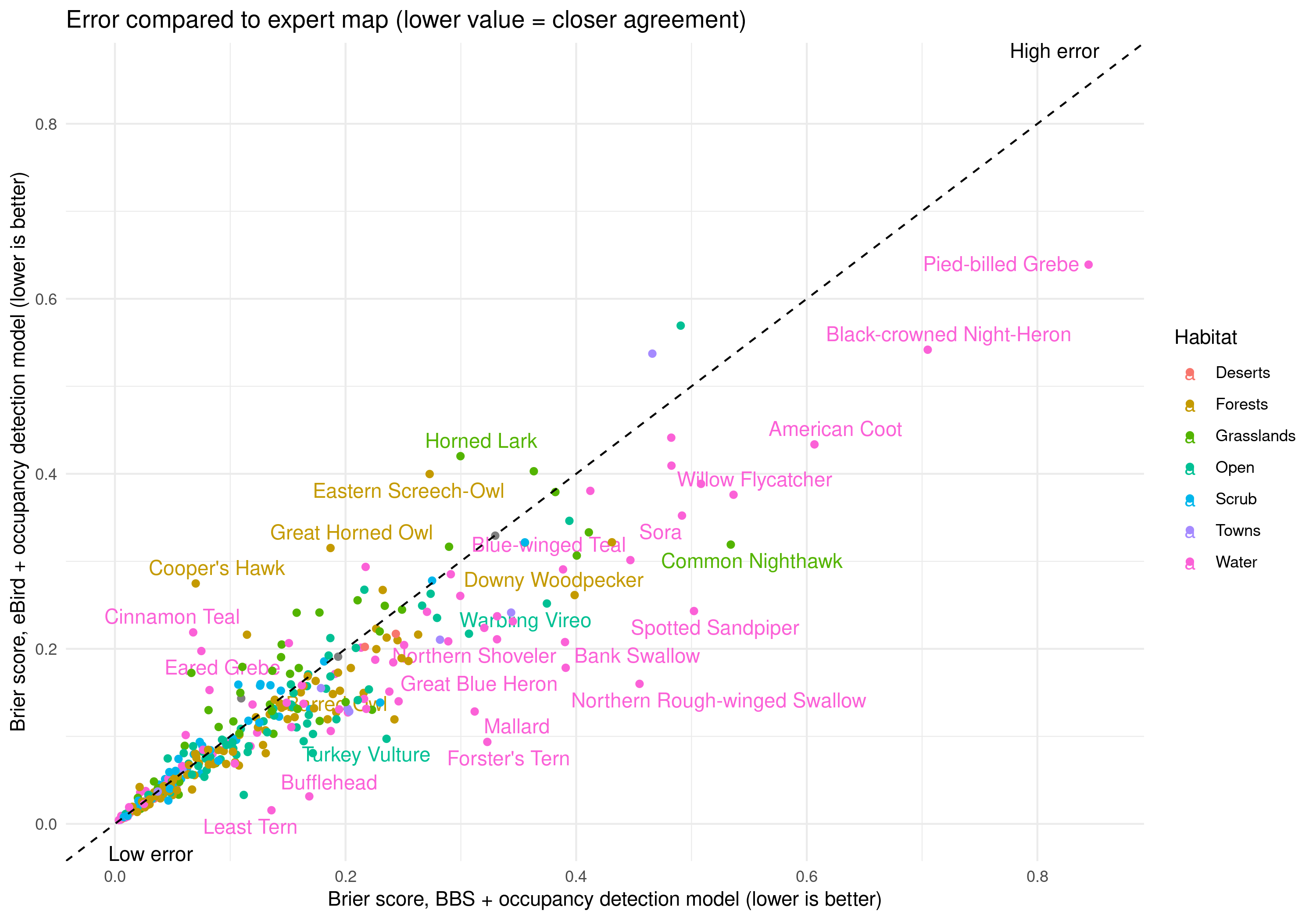}
  \caption{Scatter plot of Brier scores obtained by an MSOD model fitted to BBS
    (x-axis) against those obtained by an MSOD model fitted to eBird
    (y-axis). The points are coloured by the habitat assigned to them by the
    \href{https://www.allaboutbirds.org/news/}{All About Birds} website. The
    dashed line indicates equal error.}
  \label{fig:brier-scatter}
\end{figure}

In the main text, we describe how we use Brier scores to assess agreement with
expert range maps. The two models which performed best on this metric were
multi-species occupancy detection (MSOD) models fitted to eBird and the North
American Breeding Bird Survey (BBS). \autoref{fig:brier-scatter} compares the
Brier scores from these models for each species. They are coloured by the
habitat assigned to them by the \href{https://www.allaboutbirds.org/news/}{All
  About Birds} website, with the habitats ``Shorelines'', ``Rivers'',
``Oceans'', ``Lakes'', and ``Marshes'' grouped together into one ``Water''
habitat. The agreement is similar for most species, but a number of water birds
seem to have a lower Brier score when using the MSOD model on eBird, indicating
somewhat better agreement with the expert maps.

\section{Comparison of coefficients found by unmarked}

In this section, we compare the coefficients estimated by our proposed maximum
likelihood approach against those estimated by the \texttt{unmarked} package
\citep{unmarked}. To do this, we use the \texttt{widewt} data included in
\texttt{unmarked}. Please see the following pages for details. They are also
included as a vignette in the ROccuPy package which accompanies this
paper.\footnote{Available at \url{https://github.com/martiningram/roccupy}}

\newlength{\originalVOffset}
\newlength{\originalHOffset}
\setlength{\originalVOffset}{\voffset}   
\setlength{\originalHOffset}{\hoffset}

\setlength{\voffset}{0cm}
\setlength{\hoffset}{0cm}
\includepdf[pages=-]{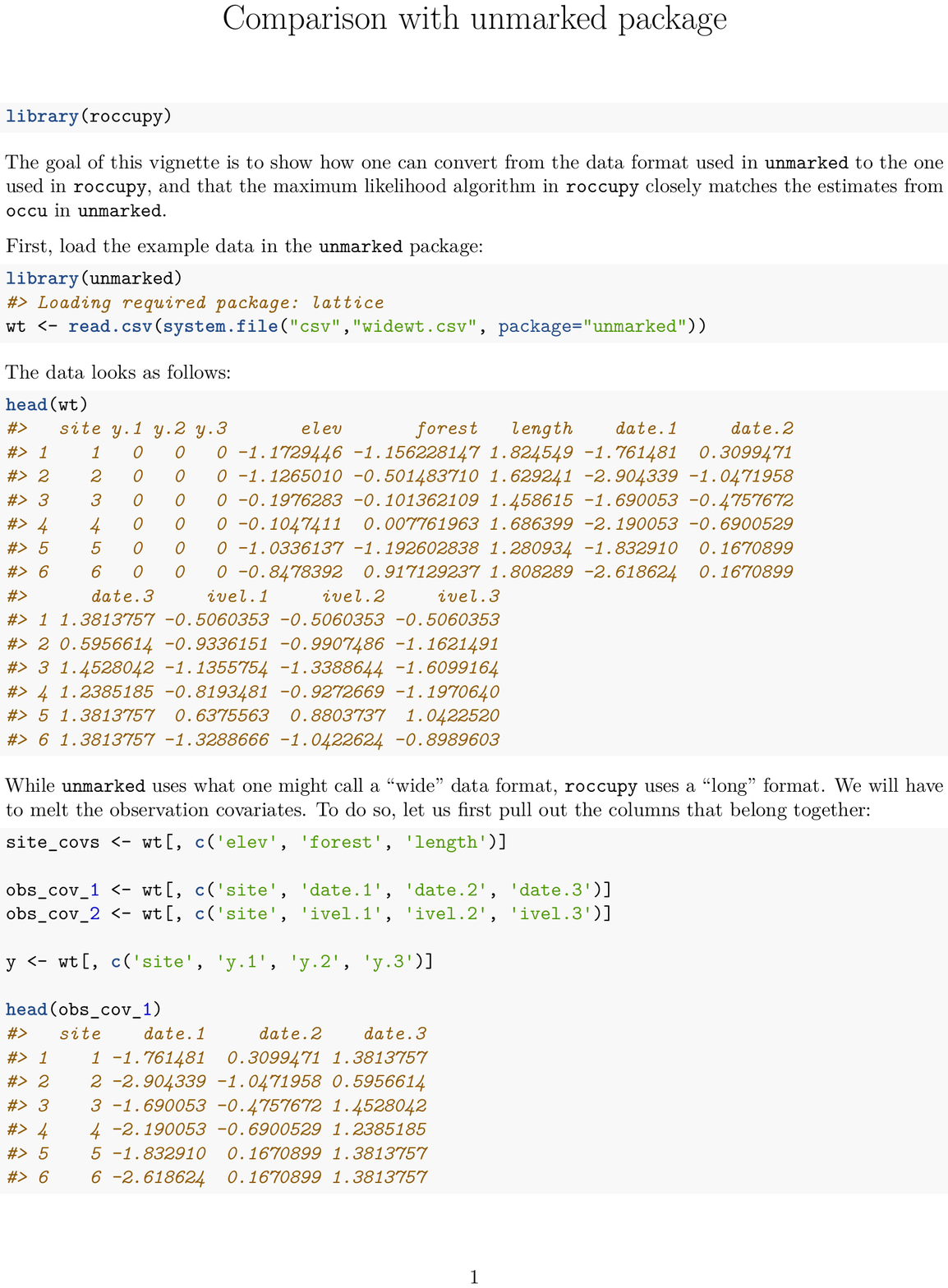}
\setlength{\voffset}{\originalVOffset}
\setlength{\hoffset}{\originalHOffset}

\section{Visualising the uncertainty in model predictions}

As mentioned in the main text, we typically produce maps of species presence
using the posterior mean probabilities. In this section, we also investigate
ways to visualise the uncertainty in the predictions.

We first clarify the different estimates that can be made. In the main text, we
discuss how the prediction of the environmental suitability of a site $i$ for
species $j$ is given by:
\begin{align}
  \textrm{logit}(\Psi_{ij}) = \bm{x}^{\text{(env)}\intercal}_i \bm{\beta}^{\text{(env)}}_j + \gamma_j.
\end{align}

Equivalently, the probability of presence $\Psi_{ij}$ can be written as:
\begin{align}
  \Psi_{ij} = \text{logit}^{-1}(\bm{x}^{\text{(env)}\intercal}_i
  \bm{\beta}^{\text{(env)}}_j + \gamma_j).
\end{align}

\begin{figure}
  \includegraphics[width=\textwidth]{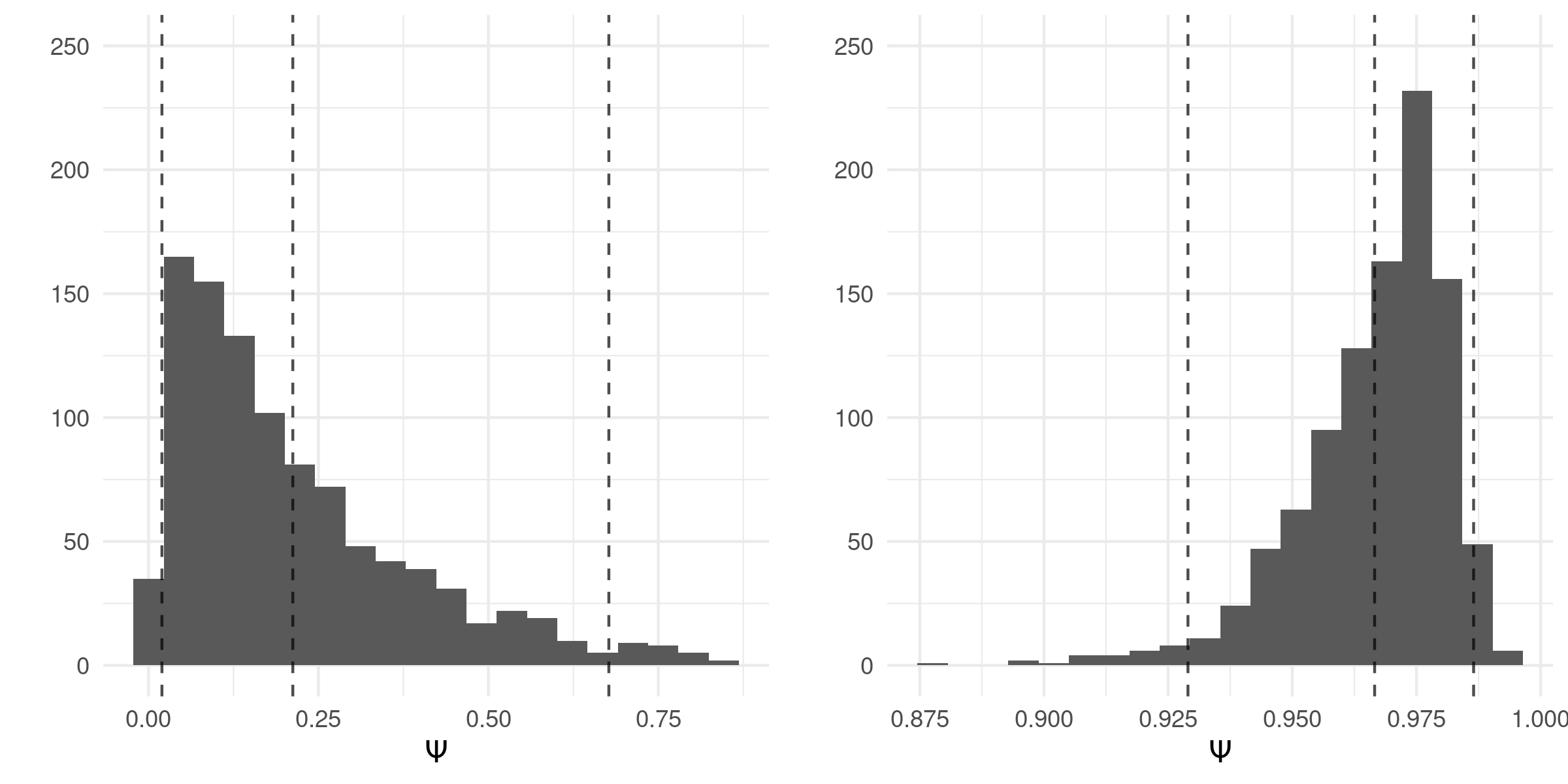}
  \caption{Two posterior distributions of $\Psi$ for different species. The two
    outer dashed lines indicate the 95\% credible interval of the posterior, and
    the inner line shows the posterior mean.}
  \label{fig:psi-posterior}
\end{figure}

Once the model has been fitted, it produces posterior distributions for
$\beta_{j}^\text{(env)}$ and $\gamma_j$. This in turn implies a posterior
distribution for $\Psi_{ij}$. Two such posterior distributions are shown in
\autoref{fig:psi-posterior}. The left panel shows a prediction with large
uncertainty: the posterior distribution includes both very low and fairly high
probabilities. The posterior mean is at around 21\%, indicating some chance that
the species is present, but the 95\% credible interval is [2.0\%, 67.7\%],
reflecting the large amount of uncertainty in the posterior. On the other hand,
the plot on the right shows a confident prediction, with a mean at 96.6\% and a
95\% credible interval of [92.9\%, 98.6\%].

\begin{figure}
  \centering
  \includegraphics[width=0.9\textwidth]{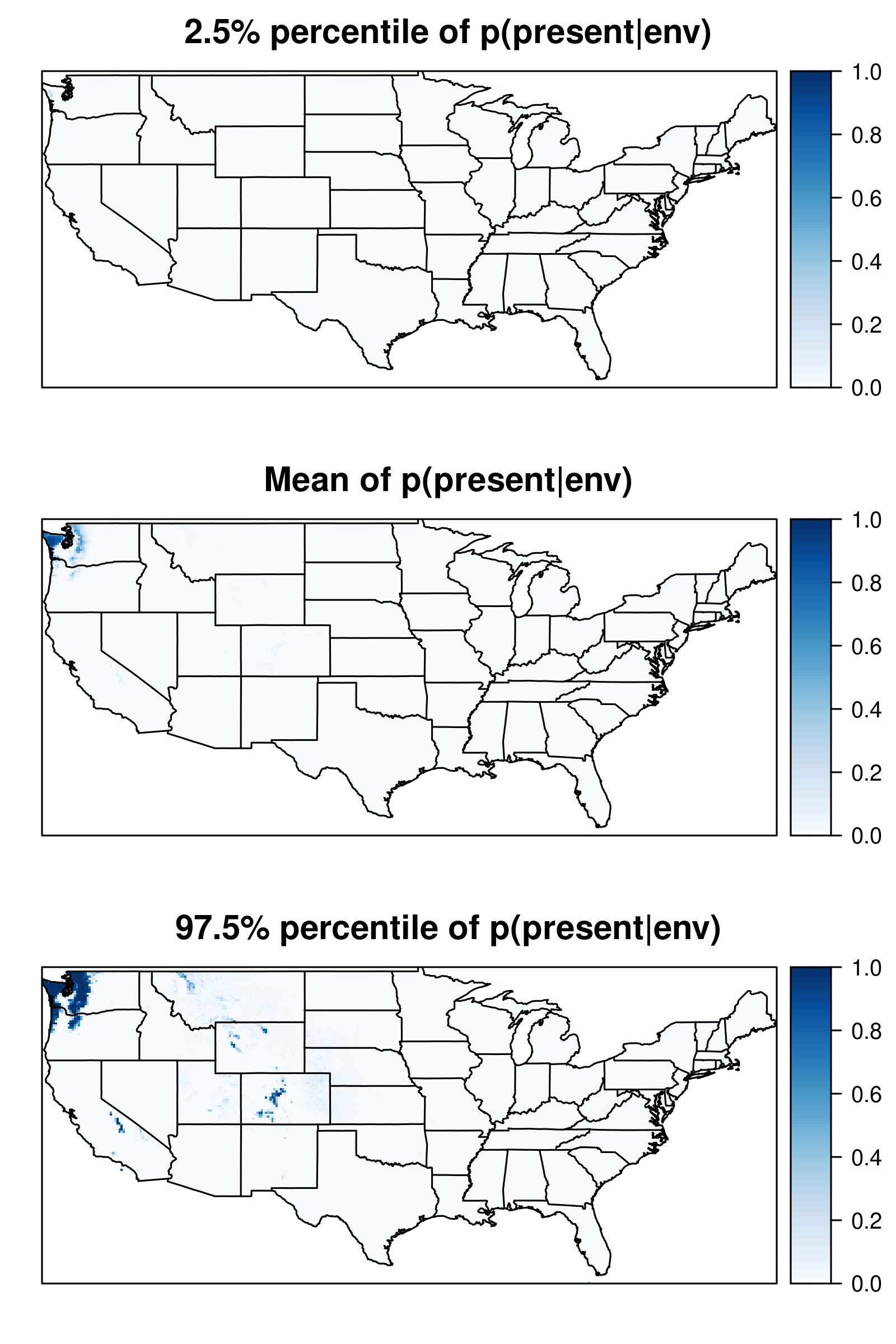}
  \caption{The three maps summarise the posterior probabilities predicted for
    \emph{Corvus caurinus}. From top to bottom, they show the 2.5\% percentile, mean, and 97.5\% percentile probabilities, respectively.}
  \label{fig:psi-interval-maps}
\end{figure}

We can show these credible intervals on a map by plotting three maps, one for
the lower end, one for the mean, and one for the upper end of the credible
intervals. This is shown in \autoref{fig:psi-interval-maps} for the species
\emph{Corvus caurinus} mentioned in the main text. The map shows that the model
is confident about which regions are \emph{not} suitable for the species, namely
almost all of the United States. There is a large amount of uncertainty in the
prediction for the northwestern corner, where the posterior probabilities range
from very low values to very high values. The model's mean prediction agrees
well with the expert map, but the figure gives a more nuanced view: the model
predicts that the northwestern corner may or may not be suitable, but that most
of the US is not suitable.

\end{appendices}

\bibliography{references.bib}
\bibliographystyle{plainnat}